\documentclass[twocolumn]{aastex61}  % twocolumn, manuscript, preprint, preprint2, modern, linenumbers, tighten
\usepackage{amsmath}

\newcommand{\Sim}{\sim\!}
\newcommand{\Approx}{\approx\!}
\newcommand{\Sersic}{S\'{e}rsic }

\shorttitle{Tidal Disruption Event Host Galaxies}
\shortauthors{Law-Smith et al.}

\begin{document}

\title{Tidal Disruption Event Host Galaxies \\ in the Context of the Local Galaxy Population}

\correspondingauthor{Jamie Law-Smith}
\email{lawsmith@ucsc.edu}

\author{Jamie Law-Smith}
\affiliation{Department of Astronomy and Astrophysics, University of California, Santa Cruz, CA 95064, USA}
\affiliation{Niels Bohr Institute, University of Copenhagen, Blegdamsvej 17, DK-2100 Copenhagen, Denmark}

\author{Enrico Ramirez-Ruiz}
\affiliation{Department of Astronomy and Astrophysics, University of California, Santa Cruz, CA 95064, USA}
\affiliation{Niels Bohr Institute, University of Copenhagen, Blegdamsvej 17, DK-2100 Copenhagen, Denmark}

\author{Sara L. Ellison}
\affiliation{Department of Physics and Astronomy, University of Victoria, Victoria, BC, Canada}

\author{Ryan J. Foley}
\affiliation{Department of Astronomy and Astrophysics, University of California, Santa Cruz, CA 95064, USA}

\begin{abstract}
We study the properties of tidal disruption event (TDE) host galaxies in the context of a catalog of $\sim$500,000 galaxies from the Sloan Digital Sky Survey. We explore whether selection effects can account for the overrepresentation of TDEs in E+A/post-starburst galaxies by creating matched galaxy samples. Accounting for possible selection effects due to black hole (BH) mass, redshift completeness, strong AGN presence, bulge colors, and surface brightness can reduce the apparent overrepresentation of TDEs in E+A host galaxies by a factor of $\sim$4 (from $\sim$$\times$100-190 to $\sim$$\times$25-48), but cannot fully explain the preference. We find that TDE host galaxies have atypical photometric properties compared to similar, ``typical''  galaxies. In particular, TDE host galaxies tend to live in or near the ``green valley'' between star-forming and passive galaxies, and have bluer bulge colors ($\Delta (g-r)\approx0.3$~mag), lower half-light surface brightnesses (by $\sim$1~mag/arcsec$^2$), higher \Sersic indices ($\Delta n_{\rm g} \approx 3$), and higher bulge-to-total-light ratios ($\Delta B/T \approx 0.5$) than galaxies with matched BH masses. We find that TDE host galaxies appear more centrally concentrated and that all have high galaxy \Sersic indices and $B/T$ fractions---on average in the top 10\% of galaxies of the same BH mass---suggesting a higher nuclear stellar density. We identify a region in \Sersic index and BH mass parameter space that contains $\sim$2\% of our reference catalog galaxies but $\ge\!60\%$ of TDE host galaxies. The unique photometric properties of TDE host galaxies may be useful for selecting candidate TDEs for spectroscopic follow-up observations in large transient surveys.
\end{abstract}


%%%%%%%%%%%%%%%%%%%%%%%%%%%%%%%%%%%%%%%%%%%%%%%%%%
\section{INTRODUCTION}\label{sec:intro}
The cores of many galaxies undergo intense nuclear activity during their lifetimes. This activity inevitably leads to the growth of the central supermassive black hole (SMBH) but is short lived compared to galactic ages and was more prevalent when the universe was only $\sim$20\% of its current age \citep{Soltan:1982, Ho:2009}. Quiescent SMBHs starved of fuel are common in the local universe \citep{Greene:2007} and are being discovered in nearby galaxies.

The presence of quiescent SMBHs in the nuclei of galaxies has been directly inferred from the dynamics of the stars and/or gas near their centers \citep[e.g.,][]{Kormendy:2013, McConnell:2013}. 
For galaxies too distant to accurately resolve the nuclear stellar or gas kinematics, it is possible to probe the presence of an SMBH with the fate of the central, closely-packed stars. 
Each star within the nuclear star cluster traces out an intricate orbit under the combined influence of the SMBH and other stars. The orbits are gradually altered owing to the cumulative effect of encounters. 
As a result, stars that are scattered into orbits that pass too close to the central SMBH can be ripped apart by the black hole's tidal field in what is known as a tidal disruption event \citep[TDE;][]{Hills:1975, Frank:1976, Rees:1988}. 
After the star is disrupted, up to half of the stellar debris falls back and accretes onto the SMBH \citep{Carter:1982, Evans:1989, Lodato:2009, Guillochon:2013}. The accretion powers a flare that is a definitive sign of the presence of an otherwise quiescent SMBH.

TDEs are identified by a combination of a rapid increase in flux, proximity to a host galaxy's nucleus, and a decay in luminosity that loosely follows the canonical $t^{-5/3}$ law, though the most compelling events are those in which the rise, peak, and decay of the transient are observed with a frequent cadence \citep[e.g.,][]{Komossa:2004,Gezari:2009,Gezari:2012,Chornock:2014,Arcavi:2014,Holoien:2014,Miller:2015}. The (well-sampled) light curves of TDEs contain vital information about the disruption and can be used to constrain the properties of the SMBH and the stellar object that was disrupted \citep[e.g.,][]{Guillochon:2014, Law-Smith:2017}. 
A few dozen candidate TDEs have been observed in the optical, UV, and X-ray \citep[for a summary, see][]{Komossa:2015, Auchettl:2017}. Future surveys such as the Large Synoptic Survey Telescope (LSST) will likely find hundreds to thousands more events \citep{van-Velzen:2011}.

The observed rates of TDEs and, in particular, the relative rates of flares in different galaxy hosts, hold important discriminatory power over both the dynamical mechanisms operating in galactic nuclei and the nature of their underlying stellar populations. However, the dynamical mechanisms that feed stars into disruptive orbits within nuclear star clusters remain highly uncertain. Stellar tidal disruption rates have typically been studied under the assumption of a spherical nuclear star cluster that feeds stars to the black hole (BH) through a two-body relaxation-driven random walk in angular momentum space \citep{Magorrian:1999, Wang:2004, Stone:2016a}. However, disks of stars and gas, if present, could feed stars to the BH at an enhanced rate through collisionless processes or secular instabilities \citep{Lightman:1977, Rauch:1998, Magorrian:1999, Madigan:2009, Madigan:2011, Merritt:2011, Vasiliev:2013, Antonini:2013}. 
A second massive body, such as an inspiraling moderately massive BH, could also induce large-angle scatterings of stars \citep{Ivanov:2005, Chen:2009}. These processes and others could result in favorable conditions for TDEs and might manifest as enhanced rates within particular galaxy hosts \citep[for a review, see][]{Alexander:2017}.

Understanding the host galaxies of TDEs is thus important; however, this understanding is in its infancy. Many uncertainties in the conditions necessary for tidal disruption will only be resolved through an understanding of the connection between TDEs and their host galaxies. This connection will hopefully become clearer with a larger sample of TDE host galaxies, but the current sample already shows hints of being highly unusual. 

TDEs appear to be observed preferentially in rare quiescent Balmer-strong galaxies \citep[also known as post-starburst or K+A galaxies, or more restrictively as E+A galaxies;][]{Arcavi:2014, French:2016}. In a sample selected from the Sloan Digital Sky Survey (SDSS), \citet{French:2016} found that a particular selection of E+A galaxies contained only 0.2\% of the sample but more than one-third of observed TDE host galaxies, implying a drastic rate enhancement. However, it is important to make a clear distinction between galaxies in which TDEs can occur and galaxies in which TDEs might be observable. We need to discern among (1) the intrinsic TDE rate based on stellar dynamics, (2) the rate of TDEs that produce luminous flares, and (3) the potential selection effects against detecting a TDE. In this paper, we seek to disentangle some of these issues.

We study the properties of TDE host galaxies in the context of a catalog of $\sim$500,000 galaxies from the SDSS. We consider first-order physical constraints and observational selection effects, and test whether they can account for the large preference for E+A/post-starburst galaxy hosts. To do this, we compare a sample of TDE host galaxies to matched control samples of galaxies in the local universe. We also find a new (photometric) observable that may be more broadly predictive of an enhanced TDE rate than E+A classification: the central light concentration, which is apparent in both galaxy \Sersic index and bulge-to-total-light ratio ($B/T$). 

We take our sample of TDEs from the catalog presented in \citet{Auchettl:2017} and compile galaxy properties from the SDSS galaxy catalogs of \citet{Brinchmann:2004}, \citet{Simard:2011}, and \citet{Mendel:2014}. We explore several key properties of TDE host galaxies, including stellar mass, BH mass, redshift, star formation rate (SFR), bulge colors, surface brightness, \Sersic index, bulge-to-total-light ratio, and galaxy asymmetry. We also compare TDE host galaxies to active galactic nuclei (AGNs) and star-forming (SF) galaxies across these observables.

This paper is organized as follows. We describe our data in Section~\ref{sec:data}. We present the uniqueness of TDE host galaxies in Section~\ref{sec:unique}. We explore selection effects in Section~\ref{sec:select}. We present a possible physical explanation for the overabundance of TDEs in E+A/post-starburst galaxies, as well as a new unique feature of all TDE host galaxies, in Section~\ref{sec:enhance}. We discuss and interpret our findings in Section~\ref{sec:discuss}. We study a few other properties of TDE host galaxies in Appendix~\ref{sec:other} and show correlations between properties in Appendix~\ref{sec:cor}.

%%%%%%%%%%%%%%%%%%%%%%%%%%%%%%%%%%%%%%%%%%%%%%%%%%
\section{DATA}\label{sec:data}
In this section, we describe our data sources as well as some conventions and definitions we use throughout the paper.

\subsection{Reference Catalog}
Our reference catalog is contained in the SDSS \citep{York:2000, Gunn:1998, Gunn:2006} DR7 \citep{Abazajian:2009} and is based on the main galaxy sample \citep{Strauss:2002}. We make use of the MPA-JHU catalogs\footnote{\href{http://wwwmpa.mpa-garching.mpg.de/SDSS/DR7}{http://wwwmpa.mpa-garching.mpg.de/SDSS/DR7}} \citep{Brinchmann:2004} of $\sim$700,000 galaxies, the \citet{Simard:2011} catalog of bulge+disk decompositions and photometry for 1.12 million galaxies, and the \citet{Mendel:2014} catalog of bulge, disk, and total stellar mass estimates for $\sim$660,000 galaxies.

We obtain redshift, bulge $g-r$, bulge and galaxy magnitude, galaxy half-light radius, galaxy \Sersic index (see Equation~\ref{eq:sersic}), bulge fraction ($B/T$), galaxy asymmetry indicator, and inclination measurements from the \citet{Simard:2011} catalog. We obtain velocity dispersion, H$\alpha$ equivalent width (EW), Lick H$\delta_A$, D$_n$(4000), and star formation rate (SFR) measurements from the MPA-JHU catalog. Here, we define H$\alpha$ EW as \texttt{H\_ALPHA\_FLUX}/\texttt{H\_ALPHA\_CONT}. We obtain total and bulge stellar masses from the \citet{Mendel:2014} catalog. We collate these measurements into a catalog of $\sim$610,000 galaxies.

We then apply the following quality control requirements. Following \citet{Scudder:2012}, we remove all galaxies with negative flux or continuum measurements as these are found to be unreliable. For consistency with the samples defined in \citet{French:2016}, we require $z > 0.01$ (to prevent severe aperture bias), reliable H$\alpha$ EWs (\texttt{H\_ALPHA\_EQW\_ERR} $>$ -1 in the MPA-JHU catalog), and median signal-to-noise ratio (S/N) per pixel of the integrated spectrum of greater than 10. Applying these selection criteria leaves us with a final catalog of~$\sim$500,000 galaxies---we refer to this as our ``reference catalog'' throughout the paper. Unless otherwise specified, our quoted errors are those provided with the measurements from their respective catalogs.

We plan to compare our sample of TDE host galaxies to AGN and SF galaxies within our reference catalog. The standard way of distinguishing AGN from SF galaxies is through the so-called BPT diagram of emission-line ratios of \citet{Baldwin:1981}; we use the empirical diagnostic of \citet{Kauffmann:2003} that is based on this approach. We classify AGN and SF galaxies as those with a minimum S/N of the four spectral lines used in the BPT classification---OIII (5007\AA), H$\beta$, NII (6584\AA), and H$\alpha$---of greater than 3. We classify low-S/N AGN and low-S/N SF galaxies as those with a minimum S/N of less than 3. 

We use the $M_\text{bh}$-$\sigma_e$ scaling from \citet{Kormendy:2013} to estimate galaxy BH masses:
\begin{equation}\label{eq:k}
\frac{M_\text{bh}}{10^9 M_\sun}=\left(0.309^{+0.037}_{-0.033}\right)\left(\frac{\sigma_e}{200 {\rm km s^{-1}}}\right)^{4.38\pm0.29}.
\end{equation}
Equation \ref{eq:k} yields $M_\text{bh}$ values with an intrinsic scatter of 0.29~dex. We use velocity dispersion measurements from the MPA-JHU catalog and perform an aperture correction to obtain the bulge/spheroidal velocity dispersion $\sigma_e$, using Equation (3) in \citet{Jorgensen:1995}:
\begin{equation}
\log \frac{\sigma_{\rm ap}}{\sigma_e} = -0.065 \log\left(\frac{R_{\rm ap}}{R_e}\right) - 0.013\left[\log\left(\frac{R_{\rm ap}}{R_e} \right)\right]^2
\end{equation}
where $R_e$ is the effective radius of the bulge or spheroid from the \citet{Simard:2011} catalog, $R_{\rm ap}$ is the aperture radius (1.5\arcsec), and $\sigma_{\rm ap}$ is the velocity dispersion measured within the aperture.
Our errors on BH mass include the error on velocity dispersion and the intrinsic scatter in the $M_\text{bh}$-$\sigma_e$ scaling, and are $\sim$0.4~dex.
Our uncertainties on BH mass are relatively large, particularly for galaxies with low velocity dispersions (a few of our TDE host galaxies have velocity dispersions near or slightly below the 70~km\,s$^{-1}$ SDSS instrumental resolution).\footnote{We note that \citet{Wevers:2017} recently published the first homogeneously measured BH masses for a complete sample of 12 optical-/UV-selected TDE host galaxies. We use the SDSS velocity dispersions, even though they are less accurate, as our goal is to use consistent metrics in comparing between TDE hosts and our reference catalog. That being said, the TDE host galaxy BH masses we match on in this work are broadly consistent with the range found by \citet{Wevers:2017}.}
However, our analysis in this work is primarily concerned with differences between properties of TDE host galaxies and our reference catalog and so does not rely on accurate determinations of BH masses---only that they are determined homogeneously in the various samples we consider. Indeed, we often control for BH mass. 
We also performed our analysis using $M_{\star, {\rm bulge}}$ to determine BH masses using the scaling from \citet{Kormendy:2013},
\begin{equation}
\frac{M_\text{bh}}{10^9 M_\sun}=\left(0.49^{+0.06}_{-0.05}\right)\left(\frac{M_{\rm bulge}}{10^{11} M_\sun}\right)^{1.16\pm0.08},
\end{equation}
and $M_{\star, {\rm bulge}}$ estimates from the \citet{Mendel:2014} catalog, as well as using different scalings for $M_\text{bh}$-$\sigma$, and our conclusions are insensitive to these choices. In fact, if we replace BH mass with $M_{\star, {\rm total}}$ throughout our analysis, our conclusions remain the same.

When we study bulge quantities, such as the bulge color, bulge fraction ($B/T$), and bulge magnitude, obtained from the \citet{Simard:2011} catalog of bulge+disk decompositions, we will show measurements from all galaxies in our reference catalog. Note that this will include galaxies where a second component is not statistically justified in the fit. We can isolate a relatively ``pure" sample of bulges by including only galaxies for which the data support a bulge+disk decomposition compared to a single \Sersic fit \citep[for example, by requiring $P_{pS} < 0.32$; see][]{Simard:2011}. The size of this ``pure'' bulge sample depends mostly on the data quality and so it can be highly incomplete. This sample includes roughly one-third of our reference catalog and only three of our TDE host galaxies (numbers 5, 7, and 8 in Table~\ref{tab:legend}). Although the data quality cannot always statistically justify the bulge+disk decomposition, bulge measurements can be applied in a consistent way to our entire sample, and we find intriguing differences between the TDE host galaxies and our reference catalog (see Section~\ref{tab:control}).

The SDSS is biased in a few ways, and this leads to some sample limitations. Most importantly, the SDSS is inherently flux limited. This limits our sample to TDEs that are fairly low $z$, but since most observed TDEs are fairly low $z$, this is not a major problem. Our approach in this paper---of not just using the entire SDSS for our comparison, but selecting matched control samples on several parameters, forcing the parameter space to be the same---should mitigate most inherent biases in our sample.

\begin{deluxetable*}{l | l | l l l l l l}
\tablecaption{TDE Host Galaxies Used in This Work\label{tab:legend}}
\tablehead{\colhead{\#} & \colhead{Event Name} & \colhead{TDE Category\tablenotemark{a}} & \colhead{Host Name} & \colhead{Host R.A.} & \colhead{Host Decl.} & \colhead{Redshift} & \colhead{Reference}}
\startdata
1 & ASASSN-14ae & Veiled & SDSS J110840.11+340552.2 & 11:08:40.116 & 34:05:52.23 & 0.0436 & \citet{Holoien:2014} \\
2 & ASASSN-14li & X-ray & SDSS J124815.23+174626.4& 12:48:15.230 & 17:46:26.45 & 0.0206 & \citet{Holoien:2016} \\
3 & PTF-09ge & Veiled & SDSS J145703.17+493640.9 & 14:57:03.18 & 49:36:40.97 & 0.064 & \citet{Arcavi:2014} \\
4 & RBS 1032 & Possible X-ray & SDSS J114726.69+494257.8 & 11:47:26.80 & 49:42:59.00 & 0.026 & \citet{Maksym:2014}\\
5 & SDSS J1323 & Likely X-ray & SDSS J132341.97+482701.3 & 13:23:41.973 & 48:27:01.26 & 0.08754 & \citet{Esquej:2007,Esquej:2008}\\
6 & SDSS J0748 & Veiled & SDSS J074820.67+471214.3 & 07:48:20.667 & 47:12:14.23 & 0.0615 & \citet{Wang:2012}\\
7 & SDSS J1342 & Veiled & SDSS J134244.41+053056.1 & 13:42:44.416 & 05:30:56.14 & 0.0366 & \citet{Wang:2012}\\
8 & SDSS J1350 & Veiled & SDSS J135001.49+291609.7 & 13:50:01.507 & 29:16:09.71 & 0.0777 & \citet{Wang:2012}\\
9 & SDSS J0952 & Veiled & SDSS J095209.56+214313.3 & 09:52:09.555 & 21:43:13.24 & 0.0789 & \citet{Komossa:2008}\\
10 & SDSS J1201 & Likely X-ray & SDSS J120136.02+300305.5 & 12:01:36.028 & 30:03:05.52 & 0.146 & \citet{Saxton:2012}\\
a & PTF-09axc & Veiled & SDSS J145313.07+221432.2 & 14:53:13.08 & 22:14:32.27 & 0.1146 & \citet{Arcavi:2014} \\
b & PTF-09djl & Veiled & SDSS J163355.97+301416.6 & 16:33:55.97 & 30:14:16.65 & 0.184 & \citet{Arcavi:2014}\\
c & PS1-10jh & Veiled & SDSS J160928.27+534023.9 & 16:09:28.28 & 53:40:23.99 & 0.1696 & \citet{Gezari:2012}\\
d & Swift J1644 & X-ray & Swift J164449.3+573451 & 16:44:49.30 & 57:34:51.00 & 0.3543 & \citet{Bloom:2011} \\
e & PTF-15af\tablenotemark{b} & NA & SDSS J084828.13+220333.4 & 08:48:28.13 & 22:03:33.4 & 0.0790 & \citet{French:2016}\\ 
  \enddata
\tablenotetext{a}{From \citet{Auchettl:2017}.}
\tablenotetext{b}{The discovery article for PTF-15af has not yet been published in the literature, but we include it here as it is included in the \citet{French:2016} sample and defines the boundary of their wF16 selection (see Figure~\ref{fig:french}).
}
\tablecomments{TDE host galaxies 1-10 are in our reference catalog, and we use hosts 1-5 in our matching analysis (see text). Host galaxies a-e are not used in our analysis, but have published H$\alpha$ EW and Lick H$\delta_A$ measurements and are shown in Figure~\ref{fig:french}.}
\end{deluxetable*}

\subsection{TDE Host Galaxies}
We use the \citet{Auchettl:2017} catalog of 71 candidate TDEs as a parent sample of TDE host galaxies. We remove candidates with only one observation and those in the {\it Not a TDE} and {\it Unknown} categories, leaving us with 42 candidate TDEs. We use the R.A., decl., and $z$ of these host galaxies to find matches in our reference catalog described above. Of the 42 candidate TDE host galaxies, 10 are in our reference catalog; the relatively low number of matches is mainly due to the redshift and magnitude limits of the various catalogs we draw from, as well as the fact that most galaxies require an SDSS spectrum for inclusion in these catalogs.
These matches and their numbering (1-10) used throughout the paper are listed in Table~\ref{tab:legend}, along with the relevant primary references.

We will use TDE host galaxies numbered 1-5 in our matching analysis of the extent of selection effects on the overrepresentation of TDEs in quiescent Balmer-strong galaxies (Section~\ref{sec:select}). 
TDE candidates 6-9 were not identified photometrically (i.e., by their light curves) but were instead proposed as TDE candidates due to their unique spectra: they are ``extreme coronal line emitters,'' and are difficult to explain as standard AGNs. Numbers 9 and 10 are not in the MPA-JHU catalog, and so do not have velocity dispersion, H$\alpha$ EW, Lick H$\delta_A$, or SFR measurements from this catalog. Additionally, No. 8 does not have a reliable velocity dispersion measurement from the MPA-JHU catalog (it is flagged with a negative error). We use the $M_{\rm bh}$-$M_{\star, {\rm bulge}}$ relation from \citet{Kormendy:2013}, using $M_{\star, {\rm bulge}}$ measurements from the \citet{Mendel:2014} catalog, to estimate BH masses for Nos. 8, 9, and 10---these BH masses do not enter into the analysis and are only used to place these TDE host galaxies on our 2D plots versus BH mass. TDE hosts 1-5 are shown with red points and histograms throughout the paper, and TDE hosts 6-10 are shown with orange points. In our 1D stacked distributions, we will show the histograms for all TDE candidate with matches in our reference catalog (1-10) with dotted black lines.
Finally, we will show H$\alpha$ EW and Lick H$\delta_A$ measurements for five TDE host galaxies (labeled a-e, see Table~\ref{tab:legend}) not in our reference catalog in Figure~\ref{fig:french}.
These last five are not used in our analysis, yet they provide additional evidence of an overrepresentation of TDEs in E+A/post-starburst galaxies. Although the small number of TDE host galaxies precludes performing detailed statistics, we are nonetheless able to draw compelling conclusions about the uniqueness of these galaxies.

The robustness of our conclusions may suffer from small numbers. Additionally, it is possible that the 10 TDE host galaxies (and ultimately the five used in our main analysis) are a special subset of TDE host galaxies and are not representative of the parent sample of 42 host galaxies. The 10 TDE host galaxies in our reference catalog are relatively low $z$ and are not particularly faint, so that they are included in SDSS, and are therefore the most well-characterized in terms of their host properties. This is a potential source of bias in the TDE host sample.

\citet{Auchettl:2017} divide their candidate events into the categories {\it X-ray TDE}, {\it Likely X-ray TDE}, {\it Possible X-ray TDE}, and {\it Veiled TDE}. We provide this classification in Table~\ref{tab:legend}, but the number of matches in each category is too small to make robust conclusions about differences in TDE host galaxies between categories. The categorization is explained in detail in \citet{Auchettl:2017} but we summarize it here. Events in the {\it X-ray TDE} category have a well-defined and trustworthy X-ray light curve. Events in the {\it Likely X-ray TDE} category have very similar properties, yet with more limited data coverage. Events in the {\it Possible X-ray TDE} category have even more limited X-ray observations. Events in the {\it Veiled TDE} category have a well-defined optical/UV light curve but no X-ray emission near the peak.

%%%%%%%%%%%%%%%%%%%%%%%%%%%%%%%%%%%%%%%%%%%%%%%%%%
\section{UNIQUENESS OF TDE HOSTS}\label{sec:unique}

Following \citet{French:2016}, we define the following selections in order to isolate quiescent Balmer-strong galaxies. We define the strong F16 (sF16) 
selection as H$\delta_A - \sigma$(H$\delta_A$) $>$ 4.0 and H$\alpha$ EW $<$ 3.0. Here, $\sigma$(H$\delta_A$) is the error in the Lick H$\delta_A$ index. H$\alpha$ EW emission is an indicator of current star formation, and so this selects for galaxies with little ongoing star formation (i.e., with specific SF rates well below the main sequence of star-forming galaxies). H$\delta_A$ absorption, from A stars, indicates star formation within the past $\Sim$ Gyr. So, sF16 galaxies have had a strong starburst in the last $\Sim$ Gyr. We define the weak F16 (wF16) selection as H$\delta_A > 1.31$ and H$\alpha$ EW $<$ 3.0. The looser cut on H$\delta_A$ means that wF16 galaxies could have several possible star formation histories. Not accounting for selection effects, 0.2\% of our reference catalog falls in the sF16 selection and 2.3\% falls in the wF16 selection. Throughout this paper, we will define galaxies in the sF16 selection as ``E+A'' galaxies---we note that it is also common to define E+A galaxies with a stricter cut on H$\delta_A$ \citep{Goto:2007}---and galaxies in either the wF16 or sF16 selections more generally as ``quiescent Balmer-strong'' galaxies.

\begin{figure}[tbp]
\epsscale{1.2}
\plotone{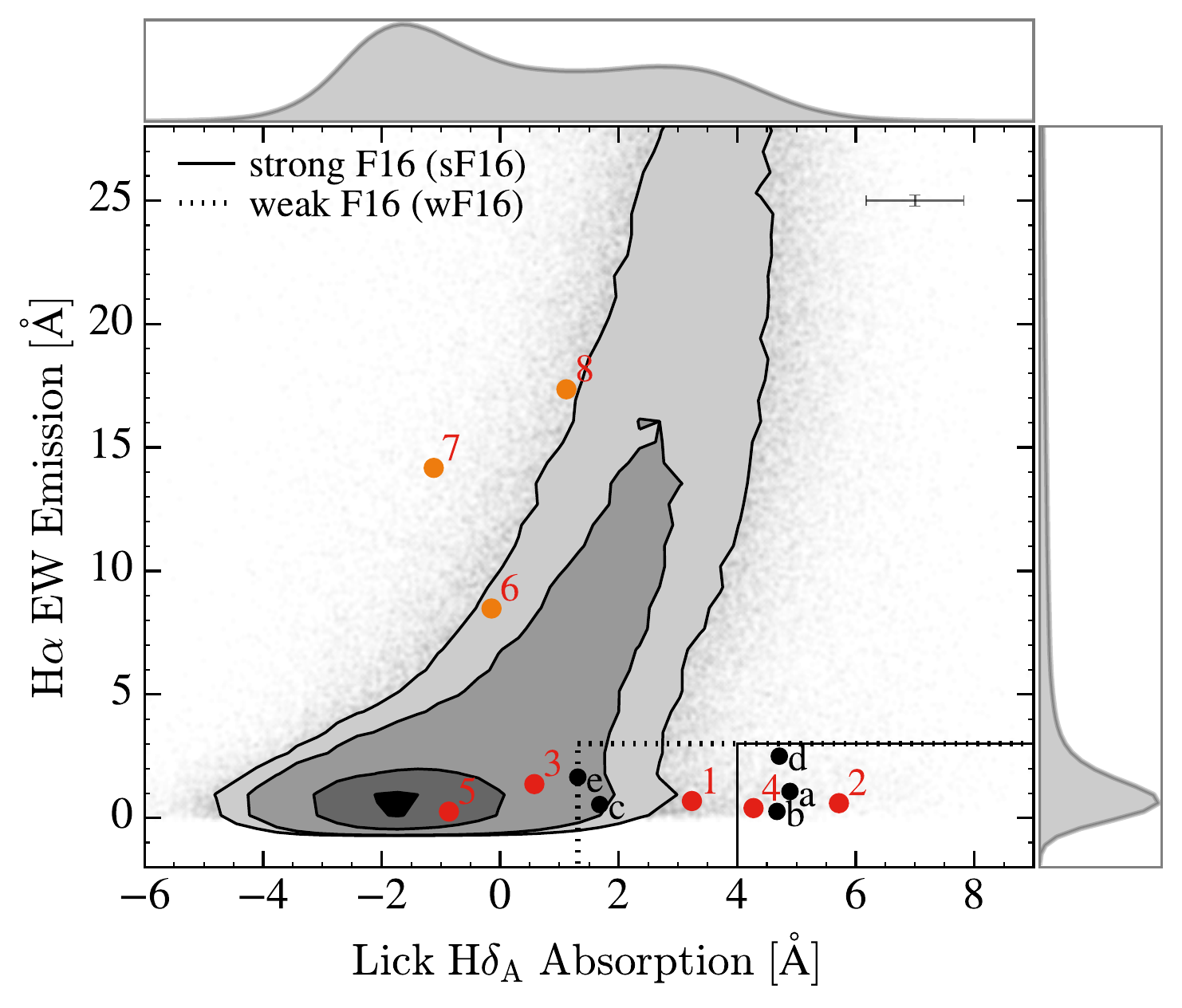}
\caption{
H$\alpha$ equivalent width emission vs. Lick H$\delta_A$ absorption, following \citet{French:2016}, for TDE host galaxies (filled circles) and our reference catalog (contours). Galaxies numbered 1-5 are used in our matching analysis (see text). The solid-line selection (including errors on Lick H$\delta_A$; sF16, see text) contains 0.2\% of the galaxies in our reference catalog and the dotted-line region (containing sF16; wF16) contains 2.3\%. Contours are spaced by $0.5\sigma$, with the darkest shading containing $0.5\sigma$ and the lightest shading containing $2\sigma$. Median errors in the TDE host galaxy measurements are shown in the top right.
}
\label{fig:french}
\end{figure}

Figure \ref{fig:french} shows H$\alpha$ EW emission versus Lick H$\delta_A$ absorption, following \citet{French:2016}, for TDE host galaxies and our reference catalog\footnote{In this and other 2D plots that follow, we use contours to show our reference catalog galaxies. Note that for a 2D distribution, $\sigma$ levels are defined differently from those for a 1D distribution. In two dimensions, the cumulative density function of a Gaussian is $F(x) = 1 - e^{-(x/\sigma)^2/2}$, meaning that ``$1\sigma$" contains 39.3\% of the volume and ``$2\sigma$" contains 86.5\% of the volume.}. TDE hosts are numbered following Table \ref{tab:legend}.
Excluding TDE candidates 6, 7, and 8 (not identified photometrically), and including candidates a-e, $3/10 = 30\%$ of the TDE host galaxies fall in the sF16 selection\footnote{Note that the sF16 selection includes the error on Lick H$\delta_A$, which excludes \#4 and d.} and $6/10=60\%$ fall in the wF16 selection.
TDEs thus remain significantly overrepresented in quiescent Balmer-strong galaxies.
A straightforward comparison to all of the galaxies in our reference catalog suggests that TDEs are overrepresented in the sF16 selection by a factor of $\sim$150 (or $\sim$190, including only the optical/UV sample defined in \citet{French:2016}) and are overrepresented in the wF16 selection by a factor of $\sim$35. 
Of the five TDE hosts we use in our matching analysis, 1/5=20\% are in the sF16 selection and 3/5=60\% are in the wF16 selection. Restricting ourselves to these five TDE host galaxies does not allow us to claim as robust an overrepresentation in sF16 galaxies,\footnote{Though the binomial false-positive percentage here for 1/5 of the galaxies in a sample being sF16 is still relatively small, $\sim$1\%; see Section~\ref{subsec:small_sample}.} but that is not the direct aim of this work. 
Our aim is to compare a wide range TDE of host galaxy properties to a larger reference catalog using consistent metrics.
Two (of four; two do not have measurements available) of the events in the \citet{Auchettl:2017} {\it X-ray TDE} category, ASASSN-14li and Swift J1644, appear to be in quiescent Balmer-strong galaxies, suggesting that X-ray TDEs share the same preference for these galaxies as do optical/UV TDEs. 

\begin{figure*}[tbp]
\epsscale{1.17}
\plotone{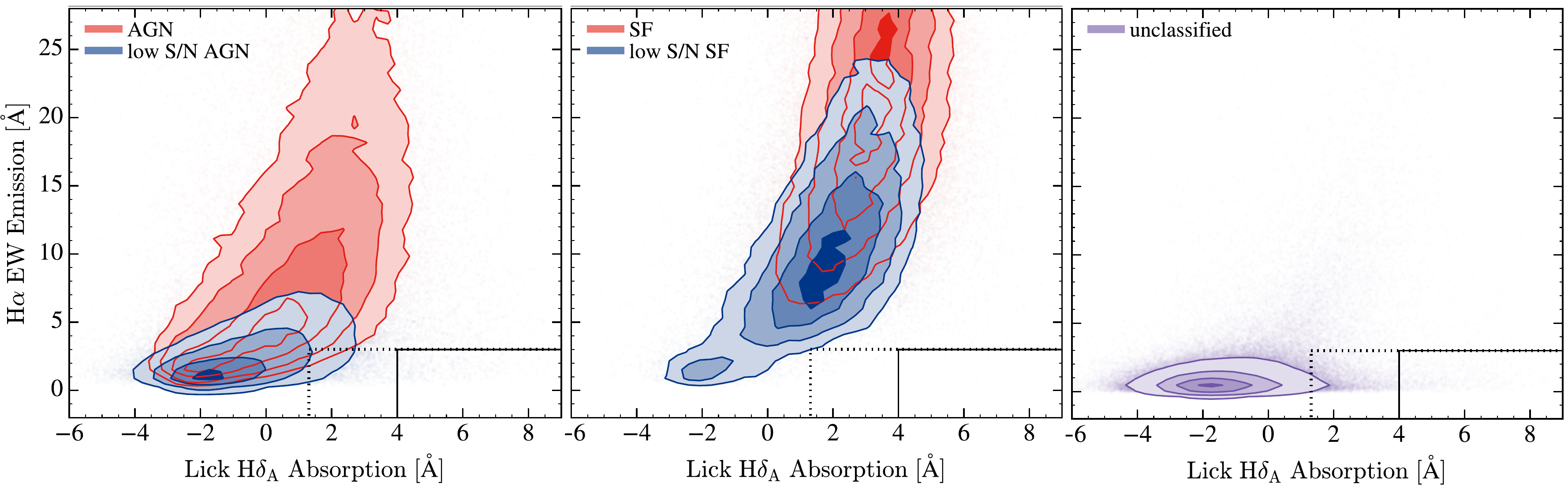}
\caption{
H$\alpha$ EW vs. Lick H$\delta_A$ as in Figure \ref{fig:french}, for our reference catalog, but split according to AGN/SF classification following \citet{Kauffmann:2003}. Low-S/N is taken as S/N $<$ 3.0. Left panel: AGN in red, low-S/N AGN in blue. Middle panel: SF galaxies in red, low-S/N SF galaxies in blue. Right panel: unclassified galaxies. Contours are spaced by $0.5\sigma$, with the darkest shading containing $0.5\sigma$ and the lightest shading containing $2\sigma$. The distributions of each subsample are normalized separately, so the relative number in each of the categories is not represented (see Table \ref{tab:AGN} for this), only their relative distributions.
}
\label{fig:french_AGN}
\end{figure*}

\begin{deluxetable}{l | lll}
\tablecaption{Fraction of Reference Catalog Galaxies in the Strong and Weak F16 Selections According to AGN/SF Classification\label{tab:AGN}}
\tablehead{\colhead{Category} & \colhead{Number} & \colhead{\% in sF16} & \colhead{\% in wF16}}
\startdata
TDE hosts (1-5) & 5 & 20 & 60 \\
TDE hosts (1-5, a-e) & 10 & 30 & 80 \\
\hline
Full reference catalog & 500,707 & 0.20 & 2.29 \\
\hline
AGN & 52,613 & 0.09 & 0.60\\
Low-S/N AGN & 93,304 & 0.36 & 4.02 \\
SF & 110,133 & 0.0 & 0.01 \\
Low-S/N SF & 42,616 & 0.01 & 0.20 \\
Unclassified & 202,041 & 0.30 & 3.61\\
\enddata
\end{deluxetable}

Next, we show where AGN and SF galaxies fall in this H$\alpha$ EW and Lick H$\delta_A$ parameter space. This is important, as there may be selection effects against detecting TDEs in some of these galaxies (in particular in galaxies hosting a strong AGN and in strongly SF galaxies), as discussed in Sections \ref{sec:select} and \ref{sec:discuss}. Figure \ref{fig:french_AGN} shows our reference catalog split into AGN, low-S/N AGN, SF, low-S/N SF, and unclassified subsamples (see Section~\ref{sec:data} for definitions). The low-S/N AGN scatter into the F16 selections more than the AGN. The unclassified galaxies also scatter into the F16 selections.

Table \ref{tab:AGN} shows the fraction of our reference catalog galaxies in the strong and weak F16 selections according to AGN/SF classification. Somewhat by construction, almost no SF or low-S/N SF galaxies are in the F16 selections. Importantly, however, if we restrict ourselves to only low-S/N AGN or unclassified galaxies, E+A/post-starburst galaxies are still rare, and TDE host galaxies remain overrepresented.

\begin{figure}[tbp]
\epsscale{1.2}
\plotone{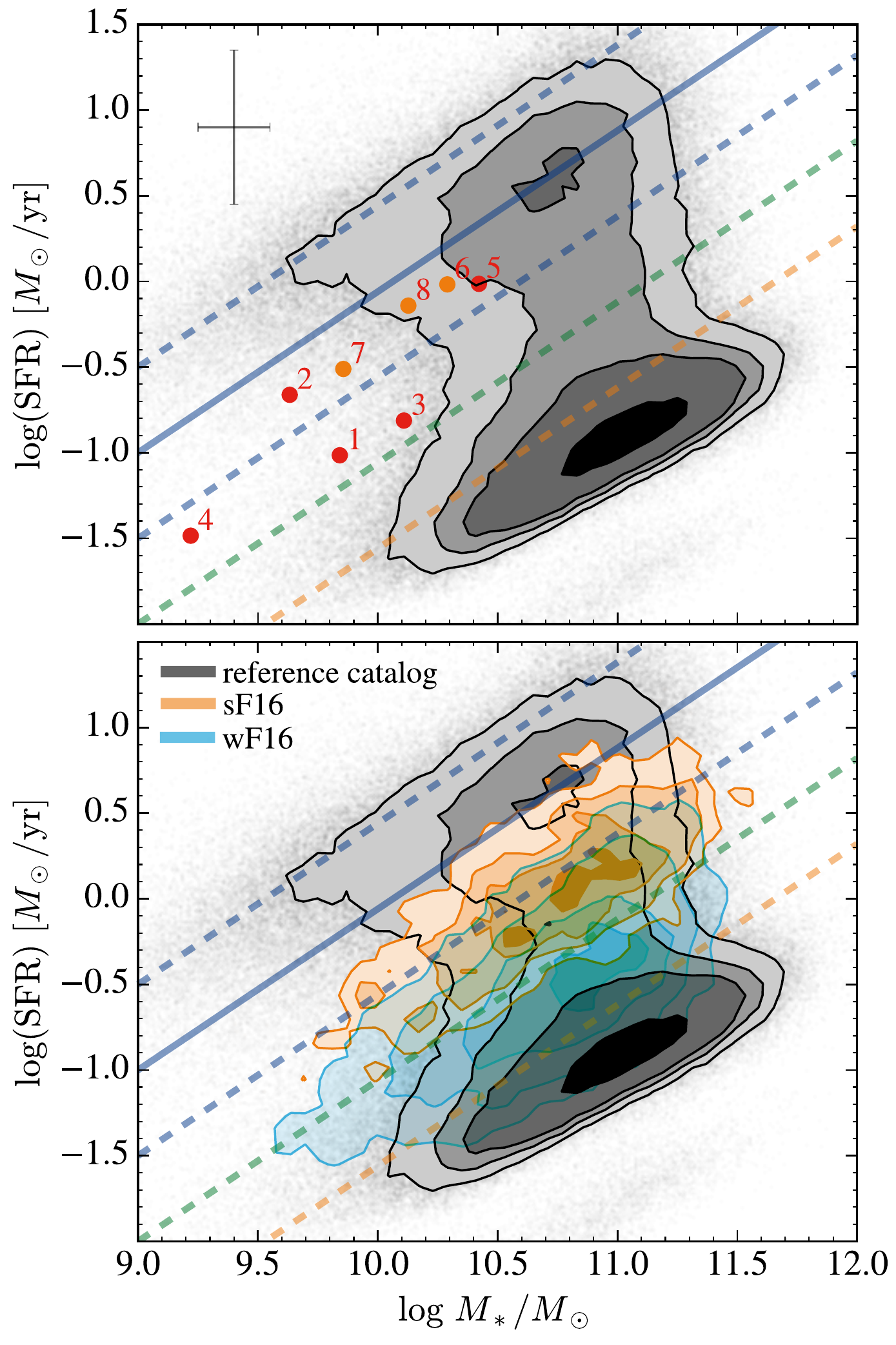}
\caption{
Top panel: total star formation rate vs. total stellar mass for TDE host galaxies (numbered points) and our reference catalog (contours). Galaxies 1-5 are used in our matching analysis. Median errors in the TDE host galaxy measurements are shown in the top left. The blue solid line describes the main sequence of SF galaxies \citep{Peng:2010}, with dashed lines spaced by 0.5~dex (the median scatter of our SFR measurements) above and below. The green and orange dashed lines are also spaced by 0.5 dex, and indicate degrees of quiescence from the SFMS. Bottom panel: the distribution for galaxies in the sF16 selection (E+A galaxies) is shown in orange and for galaxies in the wF16 selection in light blue. sF16 galaxies account for 0.2\% of our reference catalog and wF16 galaxies for 2.3\%; their distributions are normalized separately.
Contours are spaced by $0.5\sigma$, with the darkest shading containing $0.5\sigma$ and the lightest shading containing $2\sigma$. 
}
\label{fig:sfr}
\end{figure}

As a related metric of the uniqueness of TDE host galaxies, we show total star formation rate versus total stellar mass for TDE host galaxies and our reference catalog in the top panel of Figure~\ref{fig:sfr}. The solid blue line describes the star-forming main sequence \citep[SFMS;][]{Peng:2010}. We assume a $1\sigma$ scatter of 0.5~dex for the SFMS---this is the median scatter of the SFR measurements, shown by the dashed blue lines. We conservatively define the ``green valley'' or ``transition region'' in this diagram as being $1\sigma$-$3\sigma$ below the SFMS normalization \citep[e.g., see][and references therein]{Pandya:2016}---this is between the lower blue dashed line and the orange dashed line.
It is immediately apparent from Figure~\ref{fig:sfr} that none of our TDE host galaxies lie above the SFMS normalization\footnote{Though note that the errors on some of the TDE host galaxies extend above the SFMS.}. Instead, all of our TDE hosts lie below the SFMS normalization, with some being within our assumed $1\sigma$ SFMS scatter and some inhabiting the green valley.
The location of the TDE host galaxies in this diagram suggests that they could be making a transition from the SFMS toward quiescence, but additional constraints on their stellar populations and star formation histories are needed to test this hypothesis \citep[also see][]{French:2017}.

The distributions of sF16 and wF16 galaxies, separately normalized to the reference catalog's distribution, are shown in orange and light blue in the bottom panel of Figure~\ref{fig:sfr}.
We calculate the fraction of quiescent Balmer-strong galaxies in our three bands of increasing degrees of quiescence below the SFMS (each spaced by 0.5~dex).
Recall that the nominal percentage of sF16 (wF16) galaxies in our reference catalog is 0.2\% (2.3\%).
Between the solid blue line and the dashed blue line, the percentage of sF16 (wF16) galaxies is 0.5\% (0.9\%). Between the dashed blue line and the dashed green line, the percentage of sF16 (wF16) galaxies is 1.1\% (7.3\%). Between the dashed green line and the dashed orange line, the percentage of sF16 (wF16) galaxies is 0.1\% (7.8\%). 
Outside of these three bands, the percentage of sF16 and wF16 galaxies drops well below nominal.
If we restrict our reference catalog to galaxies with $\log(M_{\star,{\rm tot}}/M_\sun)<10.5$, to match the $M_{\star,{\rm tot}}$ values of our TDE hosts, the fractions of quiescent Balmer-strong galaxies quoted above inside the three bands increase only slightly.
Thus, quiescent Balmer-strong galaxies reside preferentially in the green valley, and the relative fraction of E+A/post-starburst galaxies in the band inhabited by TDE hosts 1, 3, and 4 is a factor of five greater than nominal.

%%%%%%%%%%%%%%%%%%%%%%%%%%%%%%%%%%%%%%%%%%%%%%%%%%
\section{TDE SELECTION EFFECTS}\label{sec:select}
As seen above, TDEs appear to show a distinct preference for quiescent Balmer-strong (and more restrictively E+A) galaxies. Much of this preference may in fact be due to physical and observational selection effects. In this section, we explore their extent.

\subsection{Matching}
Our strategy is to create matched comparison samples drawn from our SDSS reference catalog that are controls for the TDE host galaxies in various observables related to possible selection effects,
and then to calculate the fraction of quiescent Balmer-strong galaxies in the controls. We use TDE host galaxies numbered 1-5 for this analysis, as these TDE candidates were photometrically identified
and their hosts have measurements in our reference catalog of the properties we match. That the measurements are determined consistently between the TDE host galaxy sample and the reference catalog allows us to match and compare properties in an unbiased way.
Our results are similar if we include TDE host galaxies 6-8 (which have H$\alpha$ EW and Lick H$\delta_{\rm A}$ measurements in our reference catalog), and if we include 9 and 10 (which do not have H$\alpha$ EW and Lick H$\delta_{\rm A}$ measurements in our reference catalog).

As we match only on five TDE host galaxies, we implement our matching as simple tolerances in each parameter. We match on BH mass, redshift, bulge colors, and half-light surface brightness (motivated and discussed below).
The baseline tolerance used for matching is 1\% of the ``spread'' (=$97.5^{\rm th} - 2.5^{\rm th}$ percentile) in each parameter, which corresponds to  roughly 0.0018 in $z$, 0.037~dex in BH mass, 0.028~mag in bulge color, and 0.074~mag/arcsec$^2$ in half-light surface brightness.
We limit our control sample to a maximum of 10,000 matches per TDE host galaxy; if this is not reached, we increase the tolerance in intervals of 1\%, up to a maximum of 5\% of the ``spread'' in each parameter.
We require the same number of matches per TDE host galaxy, limited by the TDE host with the fewest matches.
We then calculate the fraction of quiescent Balmer-strong galaxies in the control. We do this matching for one parameter at a time and for several simultaneously. This allows us to control for possible selection effects in different observables without needing to understand the (likely complicated) exact form of the selection effect.
Our results are relatively insensitive to the matching technique and absolute or fractional tolerances used.

\begin{deluxetable*}{l | l | ll | ll}
\tablecaption{Fraction of quiescent Balmer-strong galaxies in control samples matched to TDE hosts 1-5. We tested all combinations of these properties, but only list combinations that (1) result in enough controls to compute a reliable fraction of sF16 or wF16 galaxies, (2) lead to an increase in these fractions, and (3) are interesting in comparison with similar combinations.
\label{tab:control}}
\tablehead{\colhead{Properties matched} & \colhead{\# Control} & \colhead{\# sF16} & \colhead{\% sF16} & \colhead{\# wF16} & \colhead{\% wF16}}
\startdata
Full reference catalog & 500707 & 996 & 0.2 & 11455 & 2.29 \\
\hline
TDE hosts (1-5) & 5 & 1 & 20 & 3 & 60 \\
TDE hosts (1-5, a-e) & 10 & 3 & 30 & 8 & 80 \\
\hline
$z$ & 50000 & 82 & 0.16 & 1089 & 2.18 \\
$M_{\rm bh}$ & 50000 & 63 & 0.13 & 1054 & 2.11 \\
bulge $g-r$ & 50000 & 1026 & {\bf 2.05} & 2635 & 5.27 \\
$\Sigma_{\rm hl,g}$ & 50000 & 117 & 0.23 & 1564 & 3.13 \\
$n_g$ & 50000 & 76 & 0.15 & 1312 & 2.62 \\
$(B/T)_g$ & 50000 & 169 & 0.34 & 1498 & 3.0 \\
\hline
$z$, bulge $g-r$ & 28600 & 227 & 0.79 & 1294 & 4.52 \\
$z$, $\Sigma_{\rm hl,g}$ & 1700 & 9 & 0.53 & 51 & 3.0 \\
$M_{\rm bh}$, bulge $g-r$ & 16680 & 81 & 0.49 & 785 & 4.71 \\
$M_{\rm bh}$, $n_g$ & 6240 & 39 & 0.63 & 496 & {\rm 7.95} \\
$M_{\rm bh}$, $(B/T)_g$ & 3655 & 12 & 0.33 & 241 & 6.59 \\
bulge $g-r$, $\Sigma_{\rm hl,g}$ & 14765 & 170 & 1.15 & 800 & 5.42 \\
bulge $g-r$, $n_g$ & 41220 & 277 & 0.67 & 2328 & 5.65 \\
bulge $g-r$, $(B/T)_g$ & 13120 & 322 & {\bf 2.45} & 1064 & {\bf 8.11} \\
$\Sigma_{\rm hl,g}$, $(B/T)_g$ & 17065 & 76 & 0.45 & 735 & 4.31 \\
\hline
$z$, $M_{\rm bh}$, bulge $g-r$ & 8025 & 40 & 0.5 & 421 & 5.25 \\
$z$, $M_{\rm bh}$, $n_g$ & 1485 & 6 & 0.4 & 129 & {\bf 8.69} \\
$z$, $M_{\rm bh}$, $(B/T)_g$ & 1560 & 4 & 0.26 & 114 & 7.31 \\
$z$, bulge $g-r$, $n_g$ & 4615 & 57 & {\bf 1.24} & 279 & 6.05 \\
$z$, bulge $g-r$, $(B/T)_g$ & 3655 & 35 & 0.96 & 180 & 4.92 \\
$z$, $\Sigma_{\rm hl,g}$, $n_g$ & 185 & 2 & 1.08 & 9 & 4.86 \\
$M_{\rm bh}$, bulge $g-r$, $\Sigma_{\rm hl,g}$ & 1780 & 7 & 0.39 & 81 & 4.55 \\
$M_{\rm bh}$, bulge $g-r$, $n_g$ & 4390 & 19 & 0.43 & 321 & 7.31 \\
$M_{\rm bh}$, bulge $g-r$, $(B/T)_g$ & 2115 & 12 & 0.57 & 164 & 7.75 \\
$M_{\rm bh}$, $\Sigma_{\rm hl,g}$, $n_g$ & 215 & 0 & 0.0 & 19 & {\bf 8.84} \\
$M_{\rm bh}$, $\Sigma_{\rm hl,g}$, $(B/T)_g$ & 580 & 3 & 0.52 & 48 & 8.28 \\
$M_{\rm bh}$, $n_g$, $(B/T)_g$ & 1580 & 10 & 0.63 & 125 & 7.91 \\
bulge $g-r$, $\Sigma_{\rm hl,g}$, $n_g$ & 3065 & 24 & 0.78 & 166 & 5.42 \\
bulge $g-r$, $\Sigma_{\rm hl,g}$, $(B/T)_g$ & 2235 & 46 & {\bf 2.06} & 161 & 7.2 \\
bulge $g-r$, $n_g$, $(B/T)_g$ & 5760 & 66 & 1.15 & 378 & 6.56 \\
\hline
$z$, $M_{\rm bh}$, bulge $g-r$, $\Sigma_{\rm hl,g}$ & 285 & 1 & 0.35 & 17 & {\bf 5.96} \\
$z$, $M_{\rm bh}$, bulge $g-r$, $n_g$ & 1105 & 4 & 0.36 & 91 & 8.24 \\
$z$, $M_{\rm bh}$, $n_g$, $(B/T)_g$ & 440 & 0 & 0.0 & 40 & {\bf 9.09} \\
$z$, bulge $g-r$, $\Sigma_{\rm hl,g}$, $n_g$ & 110 & 2 & 1.82 & 7 & 6.36 \\
$M_{\rm bh}$, bulge $g-r$, $n_g$, $(B/T)_g$ & 945 & 3 & 0.32 & 79 & 8.36 \\
\enddata
\tablecomments{$\Sigma_{\rm hl,g}$ is the $g$-band half-light surface brightness, $n_{\rm g}$ is the galaxy \Sersic index, and $(B/T)_g$ is the $g$-band bulge-to-total-light ratio. Bold numbers highlight particularly large enhancements in the fraction of sF16 or wF16 galaxies.}
\end{deluxetable*}

\begin{deluxetable*}{l | l | l l | l l}
\tablecaption{Fraction of quiescent Balmer-strong galaxies in samples created with simple cuts on the reference catalog; simultaneous matching (as in Table~\ref{tab:control}) on these combinations of parameters returns few controls. These cuts are chosen to include TDE host galaxies 1, 2, and 4 (all quiescent Balmer-strong). \label{tab:simple_cuts}}
\tablehead{\colhead{Sample} & \colhead{\# Control} & \colhead{\# sF16} & \colhead{\% sF16} & \colhead{\# wF16} & \colhead{\% wF16}}
\startdata
Full reference catalog & 500,707 & 996 & 0.20 & 11455 & 2.29 \\
\hline
TDE hosts (1-5) & 5 & 1 & 20 & 3 & 60 \\
TDE hosts (1-5, a-e) & 10 & 3 & 30 & 8 & 80 \\
\hline
Cut A\tablenotemark{a}:\\ $5.5 < \log(M_{\rm bh}/M_\sun) < 7.0$, $z<0.09$, \\ bulge $g-r<0.51$, $\Sigma_{\rm hl,g}>2.05$ & 4301 & 33 & 0.77 & 118 & 2.74 \\
\hline
Cut B: cut A plus no S/N$>$5 AGN & 4054 & 32 & 0.79 & 117 & 2.89 \\
\hline
Cut C: cut B plus $n_g > 2.24$ & 1662 & 28 & 1.68 & 85 & 5.11 \\
\hline
Cut D: cut B plus $(B/T)_g > 0.55$ & 1807 & 25 & 1.38 & 78 & 4.32 \\
\enddata
\tablecomments{$\Sigma_{\rm hl,g}$ is the $g$-band half-light surface brightness, $n_{\rm g}$ is the galaxy \Sersic index, and $(B/T)_g$ is the $g$-band bulge-to-total-light ratio.}
\tablenotetext{a}{Of the three TDE hosts (in 1-5) that pass cut A, one of three are sF16 and three of three are wF16. All of these also pass cuts B, C, and D.}
\end{deluxetable*}

\begin{deluxetable*}{l | l l l l l}
\tablecaption{Medians of 1D Distributions in Samples Matched in BH Mass to TDE Host Galaxies 1-5\label{tab:medians}}
\tablehead{\colhead{Parameter} & \colhead{Control} & \colhead{TDE hosts (1-5)} & \colhead{TDE hosts (1-10)} & \colhead{sF16} & \colhead{wF16}}
\startdata
Bulge $g-r$ [mag] & $0.78^{+1.03}_{-0.44}$  & $0.46^{+0.16}_{-0.11}$ & $0.49^{+0.32}_{-0.19}$ & $0.42^{+0.18}_{-0.08}$ & $0.67^{+0.19}_{-0.15}$ \\
$\Sigma_{\rm hl,g}$ [mag/arcsec$^2$] & $0.98^{+1.55}_{-0.68}$ & $2.06^{+2.07}_{-0.46}$ & $1.95^{+2.7}_{-0.45}$ & $1.58^{+8.28}_{-1.0}$ & $1.45^{+2.42}_{-1.04}$ \\
Galaxy \Sersic index & $1.21^{+1.29}_{-0.45}$ & $4.03^{+0.92}_{-1.55}$ & $4.3^{+1.05}_{-1.92}$ & $4.33^{+2.8}_{-1.71}$ & $2.88^{+2.58}_{-1.19}$ \\
$(B/T)_g$ & $0.06^{+0.37}_{-0.06}$ & $0.56^{+0.25}_{-0.06}$ & $0.54^{+0.26}_{-0.22}$ & $0.6^{+0.18}_{-0.25}$ & $0.44^{+0.29}_{-0.27}$ \\
$M_{g, {\rm galaxy}}$ [mag] & $-19.78^{+1.05}_{-0.9}$ & $-19.24^{+0.7}_{-0.59}$ & $-19.81^{+1.01}_{-0.49}$ & $-19.79^{+1.2}_{-0.52}$ & $-19.22^{+1.03}_{-0.76}$ \\
$M_{g, {\rm bulge}}$ [mag] & $-17.52^{+2.09}_{-1.36}$ & $-18.59^{+0.38}_{-0.35}$ & $-18.56^{+0.91}_{-0.88}$ & $-19.16^{+1.51}_{-0.43}$ & $-18.05^{+1.28}_{-1.09}$ \\
Galaxy $g-r$ [mag] & $0.56^{+0.18}_{-0.16}$ & $0.63^{+0.11}_{-0.07}$ & $0.68^{+0.09}_{-0.11}$ & $0.57^{+0.08}_{-0.07}$ & $0.74^{+0.09}_{-0.09}$ \\
Inclination & $49.18^{+21.14}_{-22.61}$ & $38.48^{+26.15}_{-22.4}$ & $40.83^{+19.56}_{-24.86}$ & $40.01^{+16.08}_{-23.61}$ & $47.69^{+22.33}_{-24.25}$ \\
\enddata
\tablecomments{$\pm$ values indicate 84th and 16th percentiles. $\Sigma_{\rm hl,g}$ is the $g$-band half-light surface brightness, $(B/T)_g$ is the $g$-band bulge-to-total-light ratio, and $M_g$ is the $g$-band absolute magnitude. Results for $r$-band are similar.}
\end{deluxetable*}

\subsection{Overview of Selection Effect Matching Results}
Table~\ref{tab:control} lists the fraction of quiescent Balmer-strong galaxies in our control samples for both individual and simultaneous matches. This table also includes results from matching on galaxy \Sersic index ($n_{\rm g}$) and bulge-to-total-light ratio ($B/T$), discussed in Section~\ref{sec:enhance}. 
We find that matching individually on redshift or BH mass slightly decreases the fraction of quiescent Balmer-strong galaxies. Matching on bulge colors increases the fraction of E+A galaxies by a factor of $\sim$10 (from 0.2\% to 2\%).
Matching on half-light surface brightness increases the fraction of quiescent Balmer-strong galaxies only slightly.
Matching on several parameters simultaneously can increase the fraction of quiescent Balmer-strong galaxies by a factor of $\sim$4. 
The number of controls within our tolerances is often too few to compute a reliable sF16 fraction when matching on more than three parameters---we address this with a simple cut-based approach later in this section, and the results of these cuts are listed in Table~\ref{tab:simple_cuts}.

Figure \ref{fig:scatters} shows 2D distributions of redshift, bulge colors, and half-light surface brightness versus BH mass as well as their 1D distributions matched on the BH masses of TDE hosts 1-5 and split into different subsamples.
This figure includes measurements for TDE hosts numbered 6-10, though the matching analysis (resulting in Table \ref{tab:control}) uses only TDE host galaxies 1-5. The 1D distributions are all smoothed and normalized to equal area. These are shown for presentation purposes only---the smoothing does not enter into or affect our analysis. 
Red distributions correspond to TDE host galaxies 1-5. We show the unsmoothed histograms for TDE hosts 1-5 in solid red and those for hosts 1-10 in dotted black. 
The different subsamples shown on the right-hand side of Figure~\ref{fig:scatters} are, from top to bottom, the TDE host galaxies, quiescent Balmer-strong galaxies (sF16 and wF16 selections), and AGNs and low-S/N AGNs. We compare to AGNs both to better understand the connection between TDEs and AGNs \citep[see, e.g.,][]{Auchettl:2017a}, and because there may be a bias against detecting TDEs in galaxies hosting a strong AGN, as discussed later in this section. 
Table~\ref{tab:medians} lists the medians and spreads of these 1D distributions matched on BH mass, as well as for some properties considered in Section~\ref{sec:enhance} and Appendix~\ref{sec:other}.

\subsection{BH Mass}
The first selection effect we consider is BH mass. As presented in the tidal disruption menu of \citet{Law-Smith:2017}, most main sequence (MS) stars cannot be disrupted outside the innermost bound circular orbit of BHs of $M_\text{bh}\gtrsim10^{7.5} M_\sun$\footnote{Though combinations of rapid BH spin and favorable orbital orientation can permit MS disruption by BHs with masses of up to a few $\times 10^8 M_\sun$ \citep{Kesden:2012}.}. Giant stars can be disrupted by higher-mass BHs, but their relative disruption rate is lower and their flares last on the order of years \citep{MacLeod:2012} and may not be seen in current surveys. BHs with $M_\text{bh} \lesssim 10^6 M_\sun$, on the other hand, may inefficiently circularize the debris from the majority of MS star disruptions \citep{Guillochon:2015}, leading to a delayed flare that is more difficult to detect due to a lower luminosity and longer timescale \citep{Hayasaki:2013, Dai:2013, Cheng:2014, Bonnerot:2016}. These lower-mass BHs can disrupt denser objects such as white dwarfs, but we do not expect this to be a significant contribution to the current sample of observed TDEs \citep{MacLeod:2016a}. 

BH mass is thus a primary physical constraint in whether a TDE can be observed in a given galaxy, and we control for it by creating matched samples with similar distributions in BH mass to the observed TDE host galaxies. We obtain BH masses from velocity dispersions, though our conclusions are insensitive to the exact method used to derive BH masses (see Section \ref{sec:data}). As long as BH mass is determined homogeneously between the different samples we consider, the effect of our uncertainty on BH mass is largely mitigated; i.e., we do not rely on accurate determinations of BH mass for our conclusions. In fact, if we simply replace BH mass with $M_{\star, {\rm tot}}$ throughout our analysis, our conclusions remain the same. Matching on only BH mass to TDE hosts 1-5 (see Table \ref{tab:control}) slightly decreases the fraction of quiescent Balmer-strong galaxies in our reference catalog.

We expect (and find; see Figure \ref{fig:scatters}, or \citet{Wevers:2017} for TDE hosts not in our reference catalog) nearly all currently observed TDEs to occur in the BH mass range of $10^{5.5}<M_\text{bh}/M_\sun<10^{7.5}$. 
TDE host galaxies thus have significantly lower BH masses than the bulk of our reference catalog. We plot each of the properties discussed below versus BH mass on the left-hand side of Figure \ref{fig:scatters}, and we control for BH mass (matching on TDE hosts 1-5) in the stacked 1D distributions on the right-hand side.

\subsection{Redshift Completeness}
The second selection effect we consider is redshift completeness. The luminosity of a tidal disruption flare depends on the stellar mass, stellar radius, BH mass, impact parameter, and circularization efficiency of the debris. For a typical TDE, however, the maximum peak luminosity does not vary by more than an order of magnitude, and most observed TDEs have peak luminosities of $10^{43}$ to $10^{44}$ erg s$^{-1}$. Additionally, most TDEs appear to be sub-Eddington or Eddington limited for their BHs \citep{Hung:2017}.

A typical TDE can thus only be observed out to a redshift that depends on the detection limits of the telescope and, especially if the flare is Eddington limited, the mass of the BH. \citet{Strubbe:2009} and \citet{Kochanek:2016} studied the dependence of TDE rates on BH mass and redshift in detail; generally, they found that detection rates decrease rapidly with redshift, but that future surveys such as LSST could be sensitive to a sizable number of TDEs at $z>1$.
TDE detectability is a strong function of redshift, and we control for this by creating samples matched on redshift to the TDE host galaxies in our sample.
Matching on only redshift (see Table \ref{tab:control}) slightly decreases the fraction of quiescent Balmer-strong galaxies in our reference catalog.

\begin{figure*}[tbp]
\epsscale{0.56}
\plotone{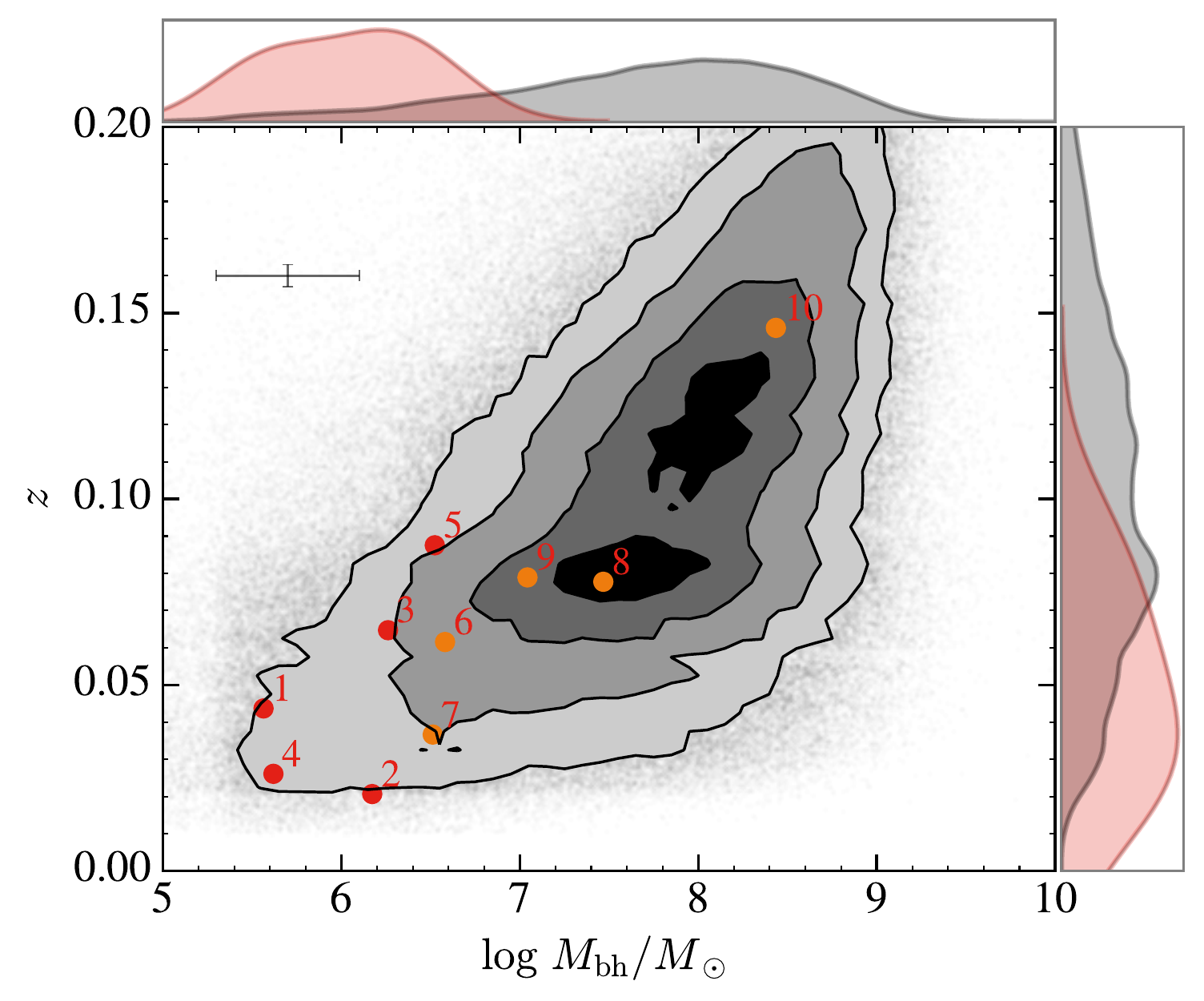} 
\plotone{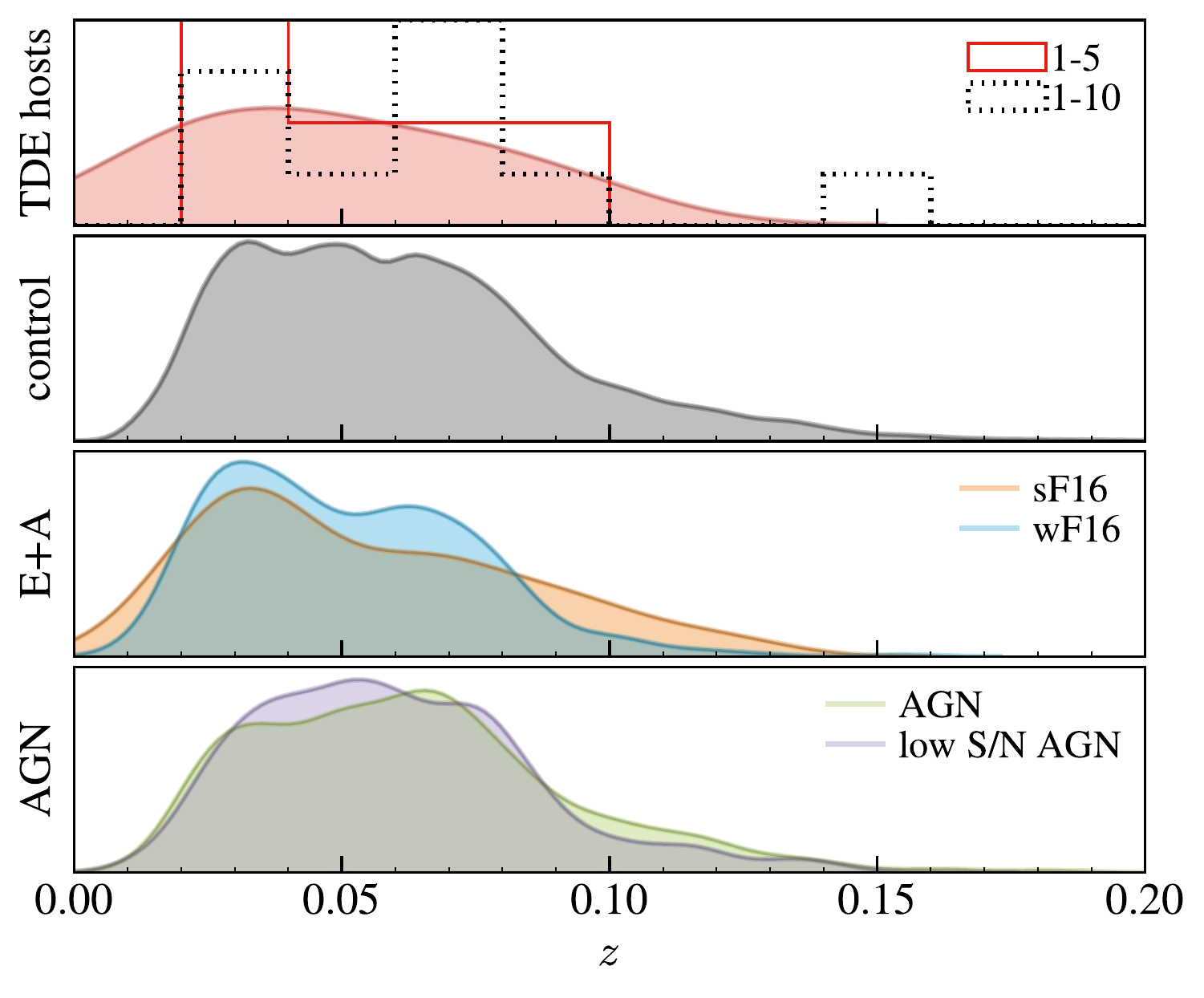}
\plotone{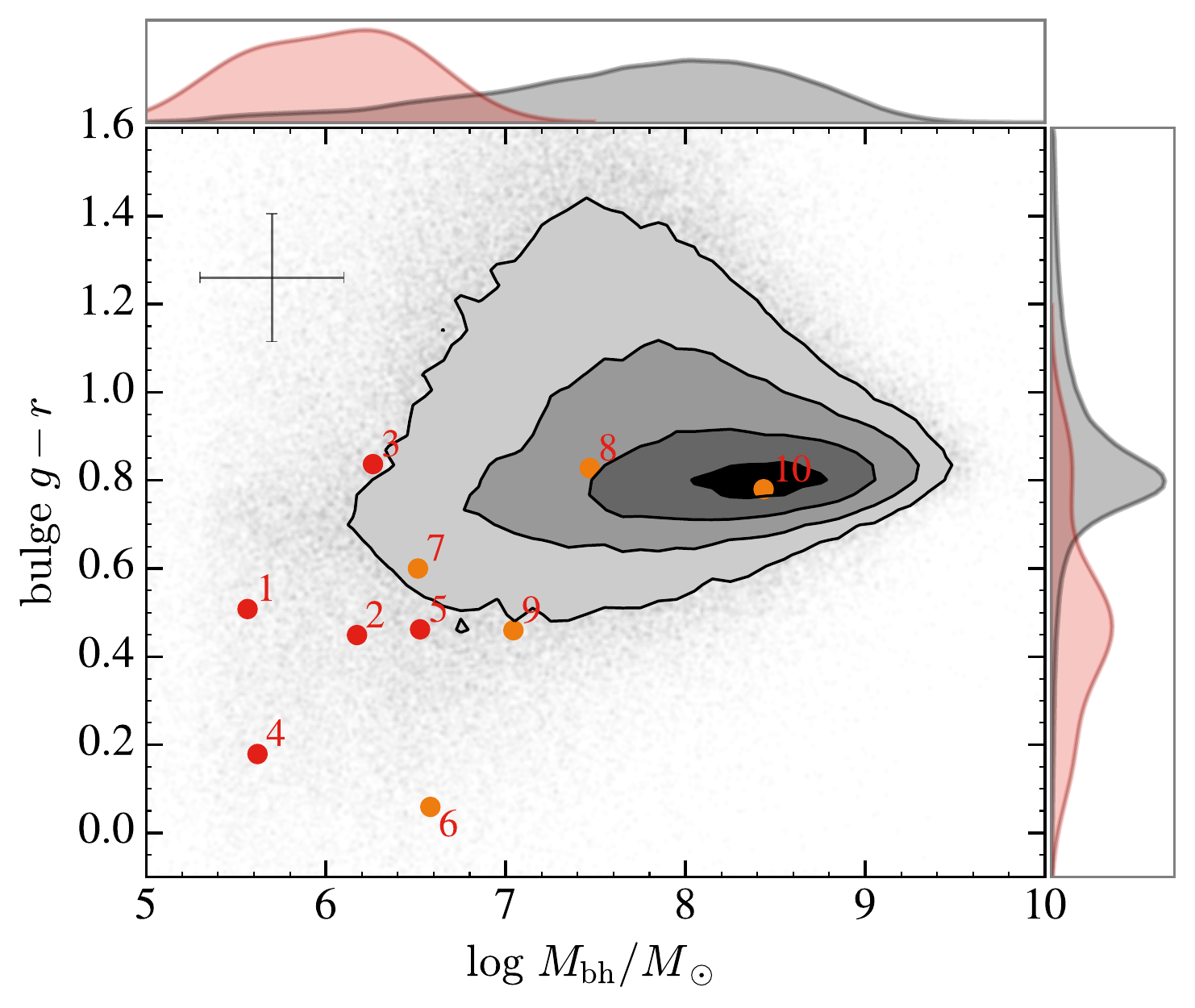} 
\plotone{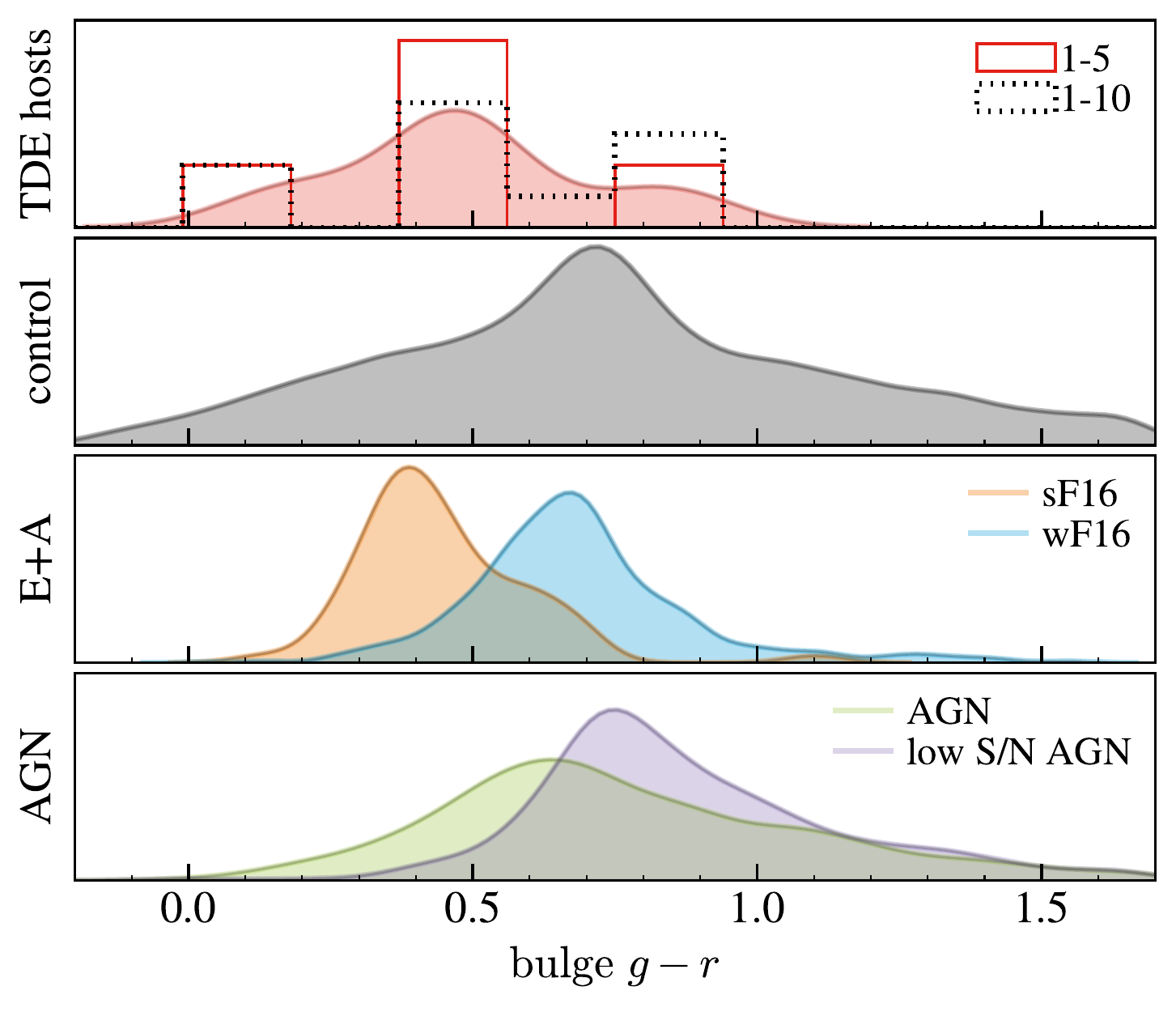}
\plotone{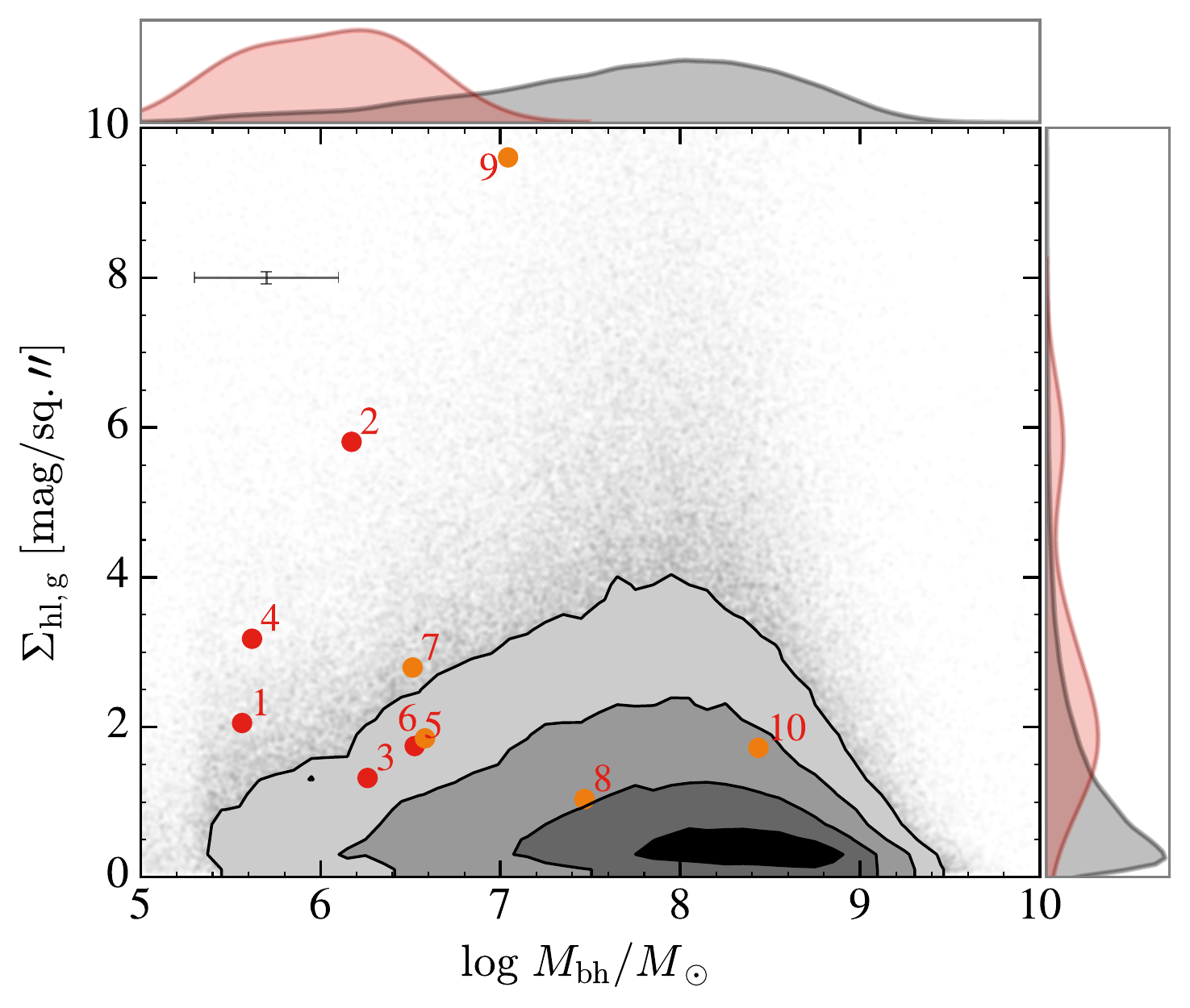} \plotone{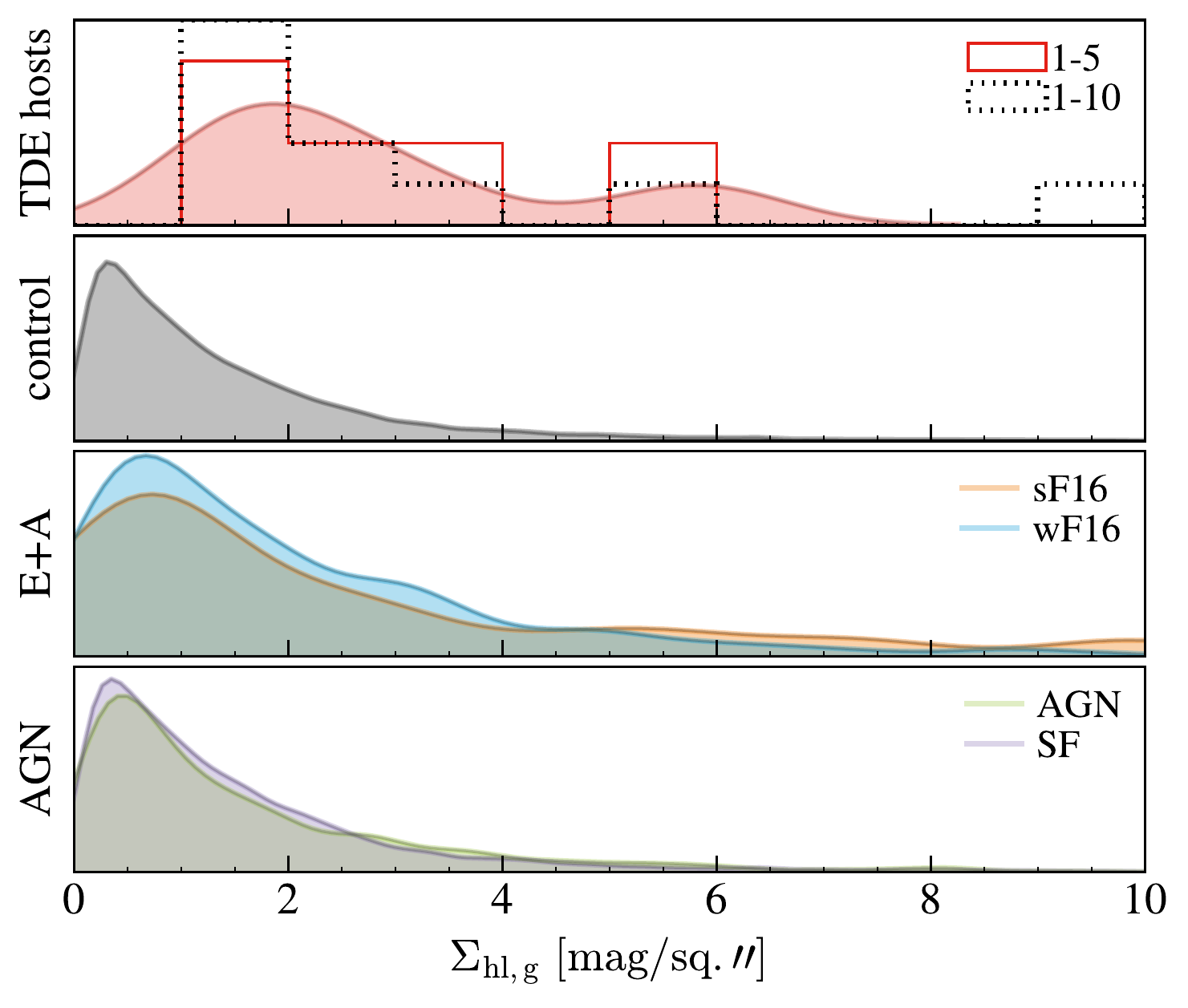}
\caption{
Left panels, top to bottom: redshift, bulge $g-r$, and half-light surface brightness vs. BH mass for TDE host galaxies (numbered points) and our reference catalog (contours). Galaxies 1-5 are used in our matching analysis. BH masses for 8, 9, and 10 are determined via $M_{\star, {\rm bulge}}$. Contours are spaced by $0.5\sigma$, with the darkest shading containing $0.5\sigma$ and the lightest shading containing $2\sigma$. Median errors in the TDE host galaxy measurements are shown in the top left. Right panels: 1D distributions in these properties in different subsamples, matched on BH mass of TDE hosts 1-5. From top to bottom in each panel, the subsamples are: TDE host galaxies (1-5 in red, showing both smoothed and actual distributions, and 1-10 in dotted black), our reference catalog (black), the strong F16 selection (orange), weak F16 selection (light blue), AGN (green), and low-S/N AGN (purple). In the bottom-right panel, we show SF galaxies, rather than low-S/N AGN, in purple, as these have a very similar distribution to the AGN. All 1D histograms are smoothed and normalized to equal area.
}
\label{fig:scatters}
\end{figure*}

We show redshift versus BH mass for TDE host galaxies and our reference catalog in the top-left panel of Figure \ref{fig:scatters}. All but one of the TDE hosts in our sample have $z<0.1$. 
The top-right panel of Figure \ref{fig:scatters} shows that, after controlling for BH mass, the redshift distributions of the TDE host galaxies, sF16 and wF16 galaxies, and both low- and high-S/N AGNs are similar to that of our reference catalog. We note that the ``sF16" classification of E+A galaxies is likely to change with redshift if it is based on a single slit width.
Although this is a small effect for galaxies with $z<0.1$, it is sizable when comparing to galaxies at $z>0.2$. The fraction of sF16 galaxies in our reference catalog as a function of redshift bin (we took $\Delta z=0.01$), for $z \lesssim 0.1$, is relatively constant at the nominal $\simeq$0.2\%, but for $z \gtrsim 0.25$, it can be $>$5\%.

\subsection{Bulge Colors}
The third selection effect we consider is the color of the galaxy bulge,\footnote{The color of the core/nucleus may be more relevant for TDE detectability but this measurement is not available for nearly as many galaxies as in our catalog drawn from SDSS.} as dusty, red bulges might obscure TDEs. 
Indeed (see below), TDE hosts have bluer bulge colors than most galaxies. We thus control for this possible selection effect by creating samples matched in bulge $g-r$ to TDE host galaxies 1-5. 
Controlling for only bulge $g-r$ (see Table \ref{tab:control}) results in a large increase in the fraction of E+A galaxies in our reference catalog: compared to the nominal percentage of 0.2\% sF16 galaxies, the matched sample has 2\% sF16 galaxies, a factor of $\sim$10 increase. The fraction of wF16 galaxies increases by a factor of two, from $2.3\%$ to $5.3\%$. Recall that $1/5=20\%$ of the TDE host galaxies used in our matching analysis (or $3/10=30\%$ including hosts not in our reference catalog) are in the sF16 selection and 3/5 (or 8/10) are in the wF16 selection.

The middle-left panel of Figure \ref{fig:scatters} shows bulge $g-r$ color versus BH mass for the TDE host galaxies and our reference catalog. In the middle-right panel (where we have controlled for BH mass), we see that TDE hosts have bluer bulge $g-r$ colors than the reference catalog,
suggesting a preference against observing TDEs in redder bulges. This is also seen clearly in the sF16 sample but only very weakly in the wF16 sample. The AGN and low-S/N AGN samples appear similar to the reference catalog, with the low-S/N AGN sample showing slightly bluer bulge colors than the AGN sample. TDE hosts 1-5 have a median bulge $g-r$ of 0.46 mag, and TDE hosts 1-10 of 0.49 mag. After matching in BH mass to TDE hosts 1-5, the reference catalog has a median bulge $g-r$ of 0.78 mag, and sF16 galaxies of 0.42 mag. This is also listed in Table~\ref{tab:medians}.
So both TDE hosts and E+A/post-starburst galaxies have median bulge colors $\sim$0.3~mag bluer than the control sample.

\subsection{Half-light Surface Brightness}
The fourth selection effect we consider is on surface brightness, as image subtraction for transients might be more challenging for high surface brightness galaxies. We define the half-light surface brightness, $\Sigma_\text{hl}$, as half the galaxy apparent magnitude divided by the galaxy half-light size in square arcseconds. Using the \citet{Simard:2011} measurements, this is $\Sigma_\text{hl,g}=(g_{g2d}+0.75254)/\pi\theta_{\rm hl}^2$, where $g_{g2d}$ is the $g$-band apparent magnitude of the GIM2D output pure \Sersic model, $g_{g2d}+0.75254$ yields half the flux, and $\theta_{\rm hl}=R_\text{chl,g}/\text{Scale}$. $R_\text{chl,g}$ is the circular half-light radius in the $g$ band, and Scale is the physical scale in kpc/arcsec$^2$ at redshift $z$.
We use only $g$ band, as results for $r$ band, or a combination of both $g$ and $r$ bands, are similar. 
Controlling for $\Sigma_\text{hl,g}$ (see Table \ref{tab:control}) increases the fraction of quiescent Balmer-strong galaxies in our reference catalog slightly, from 2.3\% to 3.1\% for wF16 galaxies.

The bottom-left panel of Figure \ref{fig:scatters} shows the half-light surface brightness versus BH mass for the TDE host galaxies and our reference catalog. It is evident  that TDEs are found preferentially in galaxies with lower half-light surface brightnesses. 
In the right panel (controlled for BH mass), we see that sF16 and wF16 galaxies have slightly lower half-light surface brightnesses than the reference catalog, and that AGN and SF\footnote{Low-S/N AGNs have a distribution in $\Sigma_\text{hl,g}$ that is indistinguishable from that of AGNs here, so we show SF galaxies instead.} galaxies have higher half-light surface brightnesses than either TDE hosts or quiescent Balmer-strong galaxies.
TDE host galaxies 1-5 have a median half-light surface brightness of 2.06~mag/arcsec$^2$, and TDE hosts 1-10 of 1.95~mag/arcsec$^2$.
After matching in BH mass to TDE hosts 1-5, the reference catalog has a median half-light surface brightness of 0.98~mag/arcsec$^2$, and sF16 galaxies of 1.58~mag/arcsec$^2$; this is also shown in Table \ref{tab:medians}. So, TDE hosts and sF16 galaxies have median half-light surface brightnesses $\sim$1~mag/arcsec$^2$ and $\sim$0.6~mag/arcsec$^2$ fainter than the control sample, respectively.

\subsection{Galaxies Hosting a Strong AGN}
We also consider a possible selection effect based on the presence of a strong AGN. Observational identification of TDEs is biased against galaxies with strong AGN, as it is difficult to distinguish a TDE signal from regular variability in a strong AGN. Galaxies with a strong AGN are often not considered for spectral follow-up on potential TDEs. As a first-order exploration of this selection effect, we performed a cut on all AGNs with S/N$>$5 (we also tried S/N$>$3) from our reference catalog. Applying this cut---both individually and in combination with controlling for other parameters---has a relatively small effect, but does slightly increase the fraction of quiescent Balmer-strong galaxies in our control samples. We do not show this high-S/N AGN cut in Table~\ref{tab:control} for clarity, as its effect is generally small, but we show its effect in combination with other simple cuts (described below) in Table~\ref{tab:simple_cuts}.

\subsection{Cumulative Effect}
Individually, controlling for BH mass or redshift slightly decreases the fraction of quiescent Balmer-strong galaxies in our control sample, controlling for half-light surface brightness slightly increases this fraction, and controlling for bulge $g-r$ increases the fraction of sF16 galaxies by a factor of $\sim$10 and of wF16 galaxies by a factor of $\sim$2.
Table~\ref{tab:control} also lists the effect of controlling for these parameters simultaneously. Note that this table also includes results from matching on two indicators of central light concentration: the galaxy \Sersic index ($n_{\rm g}$) and bulge-to-total-light ratio ($B/T$); we discuss these parameters in Section~\ref{sec:enhance}. 
We summarize our major findings with respect to matching on $M_{\rm bh}$, $z$, bulge $g-r$, and $\Sigma_{\rm hl,g}$ below.
Of these four observables, bulge $g-r$ is the most important in increasing the fraction of quiescent Balmer-strong galaxies in our control samples. However, its effect is largest on the fraction of sF16 galaxies (and remains similar for wF16 galaxies) when matched on individually. Matching simultaneously on $z$, $M_{\rm bh}$, and bulge $g-r$ increases the percentage of sF16 galaxies by a factor of 2.5, to 0.5\%. Matching simultaneously on all four parameters results in too few controls to calculate the fraction of sF16 galaxies but increases the fraction of wF16 galaxies by a factor of 2.6, to 6\%.
Recall that 1/5=20\% of the TDE host galaxies used in our matching analysis (or 3/10=30\% including galaxies not in our reference catalog) are sF16 galaxies, and 3/5=60\% (or 8/10=80\%) are wF16 galaxies.

As matching simultaneously on $z$, $M_{\rm bh}$, bulge $g-r$, and $\Sigma_\text{hl,g}$ results in too few controls to calculate the fraction of sF16 galaxies, we perform simple cuts on our full reference catalog as a cruder probe of the extent of these selection effects; this is shown in Table~\ref{tab:simple_cuts}. 
We chose cuts that are consistent with the properties of our quiescent Balmer-strong TDE host galaxies: $5.5 < \log(M_{\rm bh}/M_\sun) < 7.0$, $z<0.09$, bulge $g-r<0.51$, and $\Sigma_{\rm hl,g}>2.05$. This results in a sample with 0.77\% sF16 galaxies (a factor of $\sim$4 increase from the nominal 0.2\%) and 2.74\% wF16 galaxies (a factor of 1.2 increase from the nominal 2.3\%). Further removing all S/N$>$5 AGNs increases these numbers slightly.

In summary, the selection effects we considered in this section may reduce the apparent overrepresentation of TDEs in E+A galaxies by a factor of $\sim$4 (quoting the result from simple cuts; Table~\ref{tab:simple_cuts}) and in quiescent Balmer-strong host galaxies more generally by a factor of $\sim$2.5 (quoting the matching results; see Table~\ref{tab:control}, where we have enough wF16 galaxies to calculate a reliable fraction). 
Comparing to TDE host galaxies numbered 1-5 used in the matching, this reduces the TDE rate enhancement in sF16 galaxies to a factor of $\sim$25 (from $\sim$100) and in wF16 galaxies to a factor of $\sim$10 (from $\sim$26). Comparing to TDE host galaxies 1-5 and a-e (a-e are not in our reference catalog and so we did not match on their properties), this reduces the TDE rate enhancement in sF16 galaxies to a factor of $\sim$38 (from $\sim$150) and in wF16 galaxies to a factor of $\sim$13 (from $\sim$35).

\subsection{Small Sample Size}\label{subsec:small_sample}
As mentioned, only 1/5 of the TDE hosts we use in our matching analysis is in the sF16 selection. This is too sensitive to small number statistics to claim a true overrepresentation in sF16 galaxies using this sample alone. However, the sF16 overrepresentation increases with the larger sample of 10 photometrically identified TDE hosts with available H$\alpha$ EW and Lick H$\delta_A$ measurements, and so it appears robust. Restricting ourselves to the five TDE hosts we use in the matching, however, we can test the probability of having 1/5 sF16 galaxies (i.e., the extent to which this is due to small number statistics) in a Monte Carlo approach. We create 10,000 samples of five galaxies drawn from our full catalog and matched in $M_{\rm bh}$ and $z$ to the five TDE hosts. For each of these 10,000 samples of five galaxies, we count the number (if any) that fall in the sF16 selection. The percentage of the samples that have 1/5 or more sF16 galaxies is 1.0\%; this is the chance likelihood of having 1/5 sF16 galaxies in our sample.

\begin{figure*}[tbp]
\epsscale{0.56}
\plotone{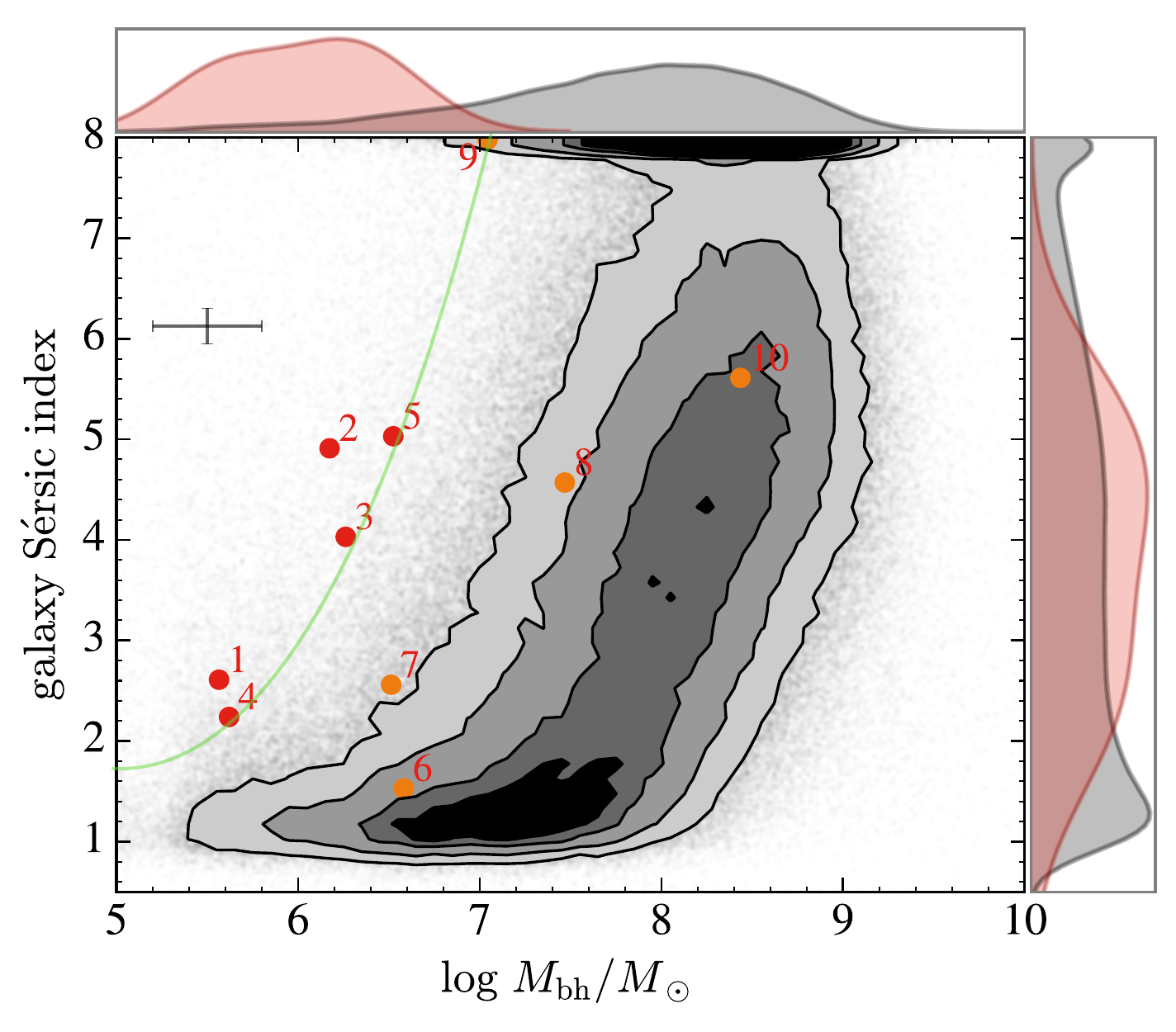}
\plotone{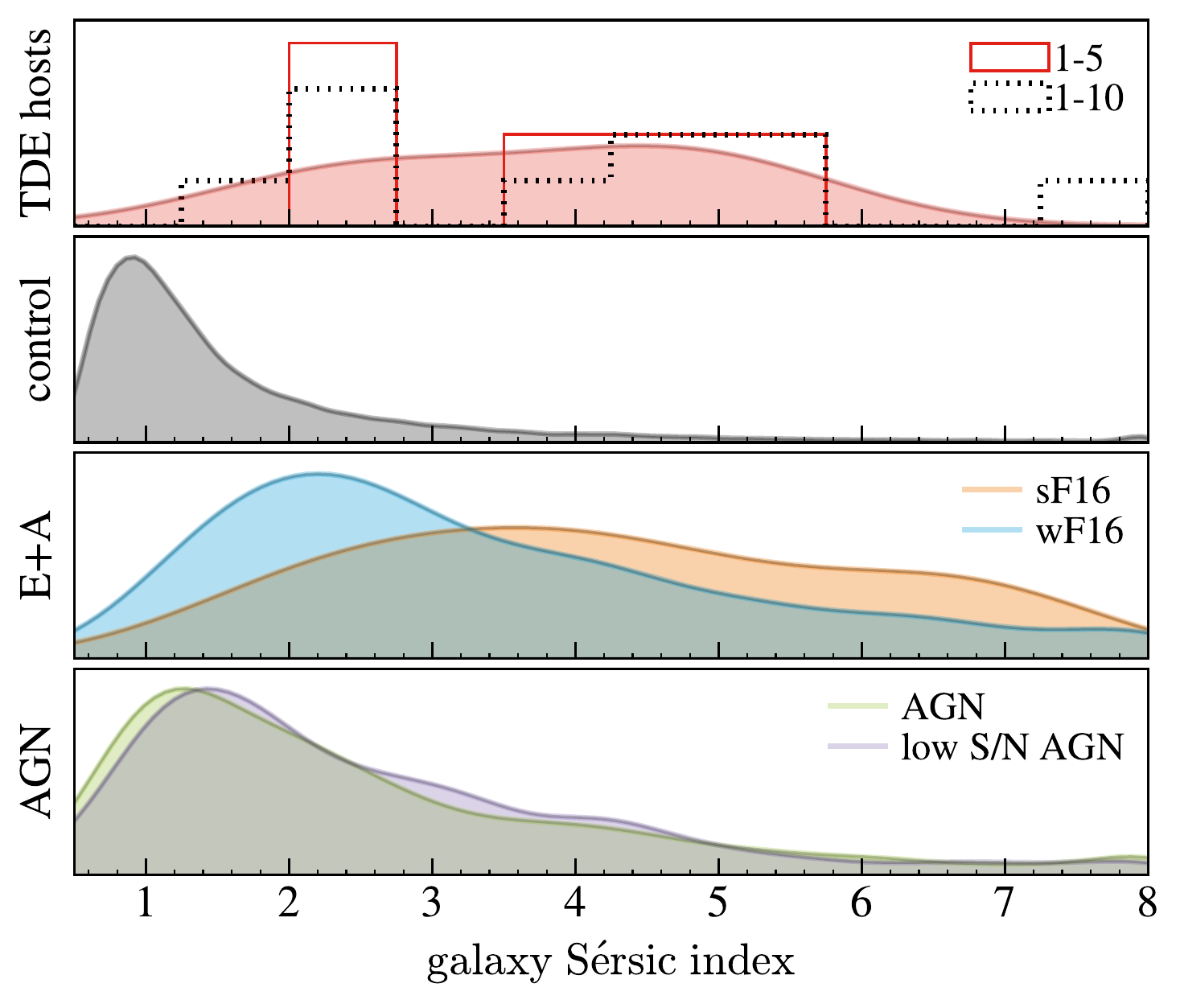}
\plotone{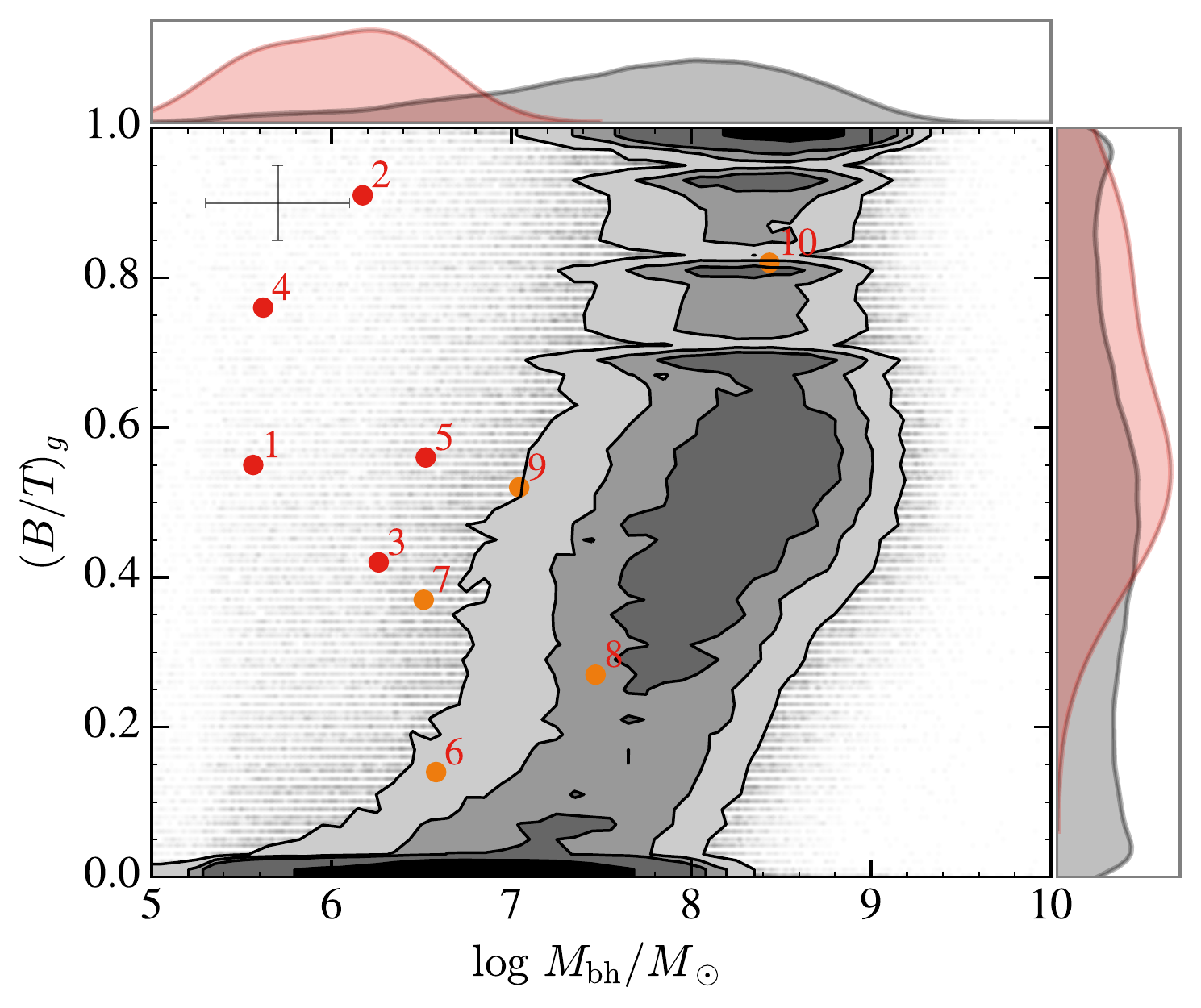}
\plotone{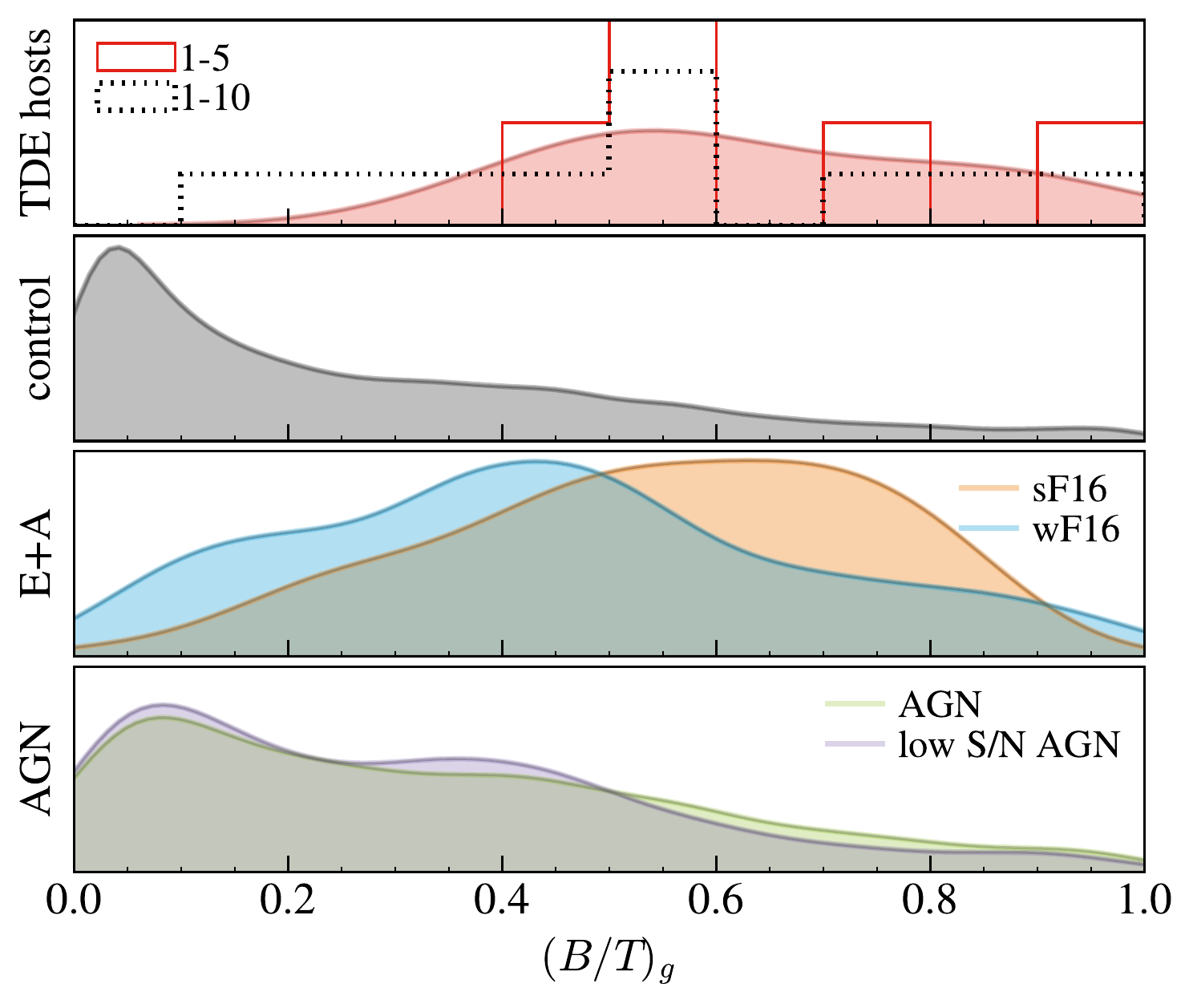}
\caption{
Top-left panel: galaxy \Sersic index vs. BH mass for TDE host galaxies and our reference catalog. We use TDE hosts 1-5 in our matching analysis. BH masses for TDE hosts 8, 9, and 10 are determined via $M_{\star, {\rm bulge}}$. Contours are spaced by $0.5\sigma$, with the darkest shading containing $0.5\sigma$ and the lightest shading containing $2\sigma$. Average errors in the TDE host galaxy measurements are shown in the top left. The region above the light green line contains $\sim$2\% of our reference catalog galaxies but 5/5 (or 6/10) of our TDE host galaxies. Top-right panel: galaxy \Sersic index distribution in different subsamples, matched on BH mass of TDE hosts 1-5. 1D histograms are smoothed and normalized to equal area. Unsmoothed 1D histograms are also shown for TDE hosts 1-5 in solid red and for TDE hosts 1-10 in dotted black.
Bottom panels: $g$-band bulge-to-total-light ratio (bulge fraction); similar description to that above. Results are similar for $r$ band.
}
\label{fig:sersic_bt}
\end{figure*}

%%%%%%%%%%%%%%%%%%%%%%%%%%%%%%%%%%%%%%%%%%%%%%%%%%
\section{PHYSICAL ENHANCEMENTS TO THE TDE RATE}\label{sec:enhance}
In this section, we consider two possible alternative (physical) explanations for the enhanced frequency of TDEs in quiescent Balmer-strong galaxies: (1) higher central stellar densities and (2) a recent merger. In exploring the first, we find a new unique property of all TDE host galaxies, regardless of E+A or quiescent Balmer-strong classification.

\subsection{Higher Central Stellar Densities}
\subsubsection{\Sersic Index}
It is expected that a higher stellar density in the nuclear star cluster surrounding an SMBH leads to a higher tidal disruption rate, as there are more dynamical encounters between stars and therefore more scatterings into the loss cone \citep[e.g.,][]{Magorrian:1999}. The galaxy \Sersic index is a broad indicator of the steepness of a galaxy's light profile and thus (to a certain extent) its stellar density profile. A \Sersic profile has the form
\begin{equation}\label{eq:sersic}
\ln I(R)=\ln I_0 - kR^{1/n},
\end{equation}
where $I$ is the intensity, $I_0$ is the intensity at $R=0$, $k$ is a constant, and $n$ is the \Sersic index. The higher the \Sersic index, the more centrally concentrated the galaxy's light profile. 

The top panel of Figure~\ref{fig:sersic_bt} shows the galaxy \Sersic index\footnote{This is for a single \Sersic fit, also referred to as $n_{\rm g}$ throughout the paper. We show galaxy rather than bulge \Sersic index here, as only a fraction of our TDE host galaxies and the reference catalog have high enough resolution data to justify (see Section \ref{sec:data}) free-$n$ \Sersic fits to their bulges. We note, however, that in this justified free-$n_{\rm b}$ sample, \citet{Simard:2011} find that galaxies with low and high $n_{\rm b}$ values also have low and high $n_{\rm g}$ values.} versus BH mass for TDE hosts and our reference catalog. Note that the \Sersic index fits are allowed to vary from 0.5 to 8 \citep[see][]{Simard:2011}. The region above the light green line contains $\sim$2\% of our reference catalog galaxies but 5/5 of TDE host galaxies 1-5 (or 6/10 including the ``extreme coronal line emitters'' 6-9 and 10 with a BH mass determined via $M_{\star,{\rm bulge}}$)\footnote{Note that this region is drawn to include these TDE host galaxies.}. We compute the fraction of reference catalog galaxies that have a higher \Sersic index than each TDE host galaxy at its BH mass (in a bin of width 0.02~dex); we find that all of the TDE host galaxies have high \Sersic indices for their BH masses. On average, TDE hosts 1-5 have galaxy \Sersic indices in the top 10\% of those of reference catalog galaxies at their BH masses. Including TDE host galaxies 6-10 results in an average \Sersic index in the top 15\%. 

In the top-right panel of Figure \ref{fig:sersic_bt}, we create a sample matched on the BH masses of TDE hosts 1-5 and compare distributions between different subsamples. TDE hosts have a much broader (toward higher values) distribution of \Sersic indices than the control sample. The distribution for sF16 galaxies is similarly weighted toward high \Sersic indices, and wF16 galaxies show a similar but weaker effect. AGNs and low-S/N AGNs show a fairly similar distribution to the reference catalog, though with a slight preference for higher \Sersic indices. TDE host galaxies 1-5 have a median galaxy \Sersic index of 4.03, and TDE hosts 1-10 of 4.30. After controlling for BH mass, our reference catalog has a median galaxy \Sersic index of 1.21, and sF16 galaxies of 4.33. This is also listed in Table \ref{tab:medians}. So both TDE host galaxies and E+A/post-starburst galaxies have relatively high galaxy \Sersic indices, especially after controlling for BH mass.

Importantly, all of our TDE host galaxies, regardless of E+A or quiescent Balmer-strong classification, have high galaxy \Sersic indices for their BH masses. We have thus identified a photometric criterion (\Sersic index) that may predict an enhanced TDE rate more broadly than a spectroscopic criterion (E+A classification).

\subsubsection{Bulge-to-total-light Ratio}
The relatively high central concentration of light in TDE host galaxies is also apparent in their bulge-to-total-light ratios. We show the $g$-band bulge fraction, $(B/T)_g$, in the bottom panel of Figure~\ref{fig:sersic_bt}. Results are similar for $r$ band. Similarly to above, we compute the fraction of reference catalog galaxies that have a higher $(B/T)_g$ than each TDE host galaxy at its BH mass. Again, we find that all of the TDE host galaxies have high bulge fractions for their BH masses: on average in the top 10\% for TDE host galaxies 1-5, and in the top 20\% for TDE hosts 1-10. Controlling for BH mass (right panel), we see that both TDE host galaxies and quiescent Balmer-strong galaxies have significantly higher bulge fractions than our reference catalog. This is related to our galaxy \Sersic index result: observed TDEs show a preference for centrally concentrated, bulge-dominated galaxies (and/or tend to avoid disk-dominated galaxies). The bulge fraction and the \Sersic index are correlated---see Figure~14 of \citet{Simard:2011} or our Figure~\ref{fig:corner}.
TDE host galaxies 1-5 have a median $(B/T)_g$ of 0.56 (0.54 for TDE hosts 1-10). Matched on BH mass to TDE hosts 1-5, the reference catalog has a median $(B/T)_g$ of 0.06, and sF16 galaxies of 0.60. This is also listed in Table~\ref{tab:medians}.

\subsubsection{Can \Sersic Index or B/T Explain the Quiescent Balmer-strong Overrepresentation?}
We perform matches on galaxy \Sersic index and bulge fraction, as we did for the selection effects considered in Section~\ref{sec:select}, and compute the fraction of quiescent Balmer-strong galaxies in the matched samples. The results of this matching (using TDE hosts 1-5 as before) are listed in Table~\ref{tab:control}; we will summarize below. We tested all combinations of the properties we studied, but only list combinations that (1) result in enough controls to compute a reliable fraction of sF16 or wF16 galaxies, (2) lead to an increase in these fractions, and (3) are interesting in comparison with similar combinations.
Matching only on \Sersic index slightly decreases the fraction of sF16 galaxies in our control sample, while matching only on $(B/T)_g$ increases the fraction of sF16 and wF16 galaxies by a factor of $\sim$1.5. Matching simultaneously on $n_{\rm g}$ or $B/T$ in combination with the selection effects we considered in Section~\ref{sec:select} can increase the fraction of quiescent Balmer-strong galaxies in our reference catalog by a factor of 3-10 depending on the combination. The highest increase in the fraction of sF16 galaxies---from 0.2\% to 2.45\%---is given by matching on bulge $g-r$ and $(B/T)_g$. The highest increase in the fraction of wF16 galaxies---from 2.3\% to 9.1\%---is given by matching on $z$, $M_{\rm bh}$, $n_{\rm g}$, and $(B/T)_g$. 

Simultaneously matching on either \Sersic index or $B/T$ and on all of the four parameters we considered earlier returns too few controls to compute the fraction of quiescent Balmer-strong galaxies. We perform simple cuts on our full reference catalog as a cruder probe of the effect of controlling for five of these properties; these are listed in Table~\ref{tab:simple_cuts}. We chose cuts that are consistent with the properties of our quiescent Balmer-strong TDE host galaxies: $5.5 < \log(M_{\rm bh}/M_\sun) < 7.0$, $z<0.09$, bulge $g-r<0.51$, $\Sigma_{\rm hl,g}>2.05$, and $n_{\rm g}>2.24$ and/or $(B/T)_g>0.55$. These result in samples with $\sim$1.7\% sF16 galaxies (a factor of $\sim$8 increase from the nominal 0.2\%) and $\sim$5.1\% wF16 galaxies (a factor of 2.2 increase from the nominal 2.3\%). 

Higher galaxy \Sersic indices and/or higher bulge-to-total-light ratios could therefore partially explain the enhanced TDE rate in quiescent Balmer-strong galaxies; we discuss this further in Section~\ref{sec:discuss}.

\begin{figure*}[tbp]
\epsscale{0.38}
\plotone{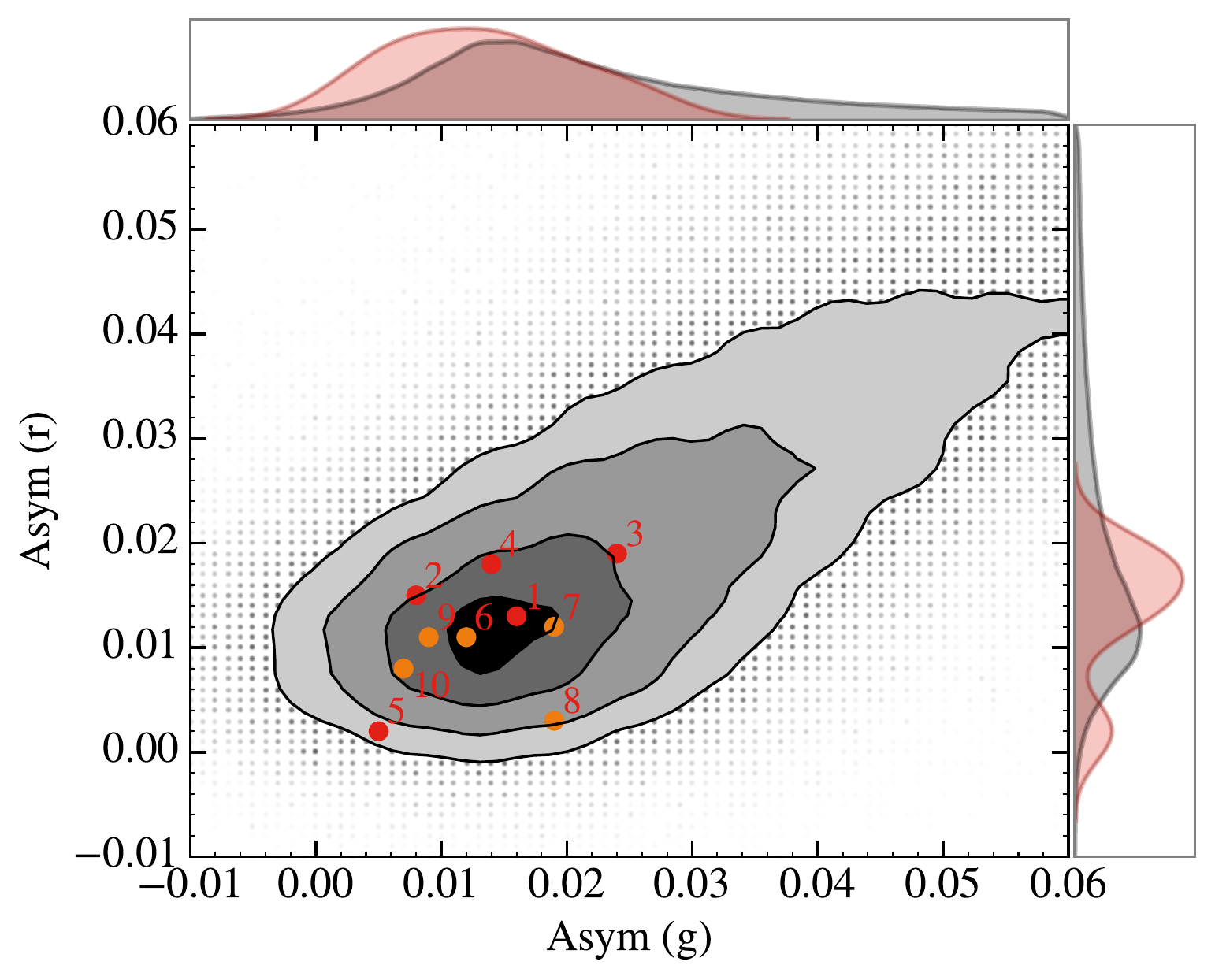}
\plotone{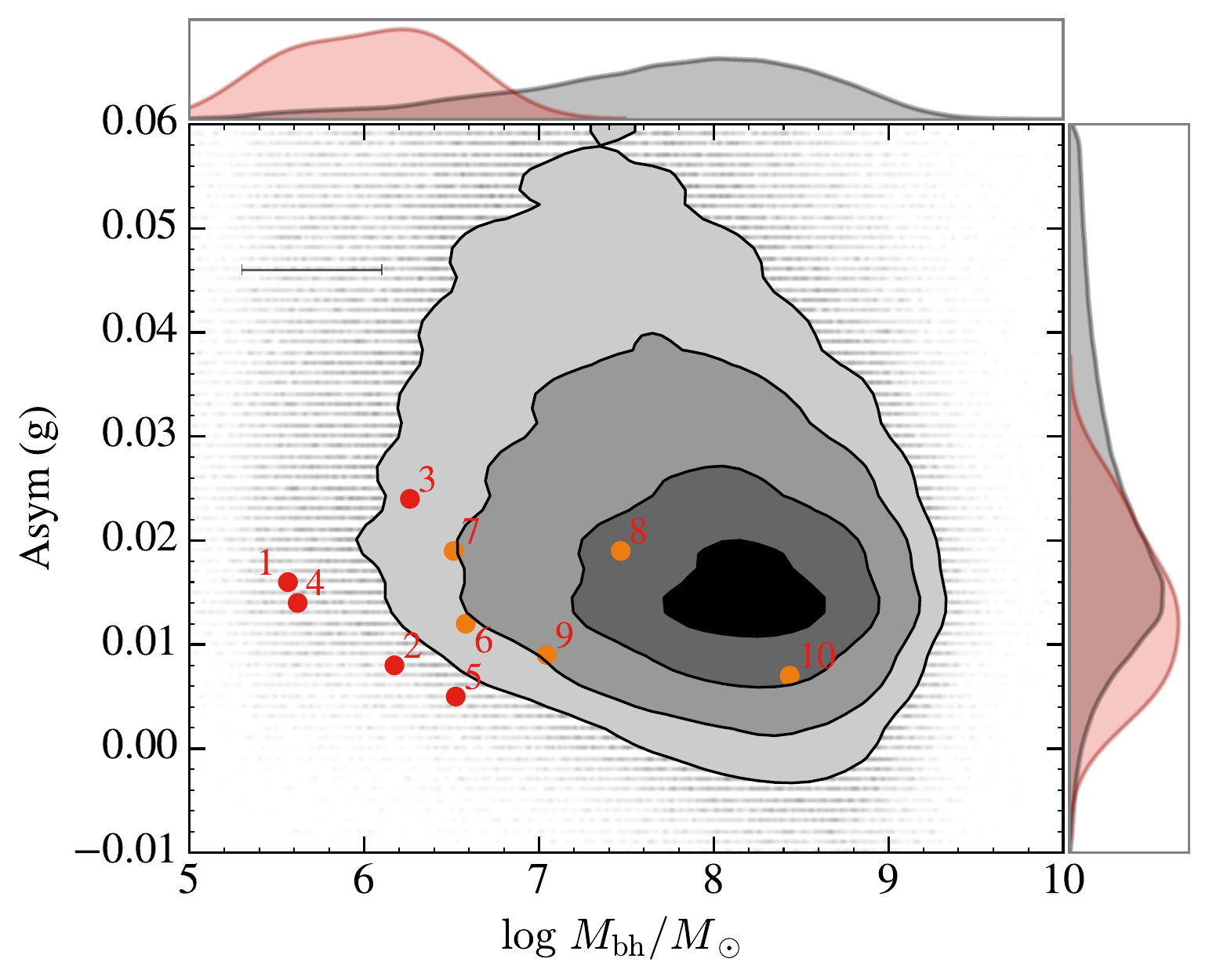}
\plotone{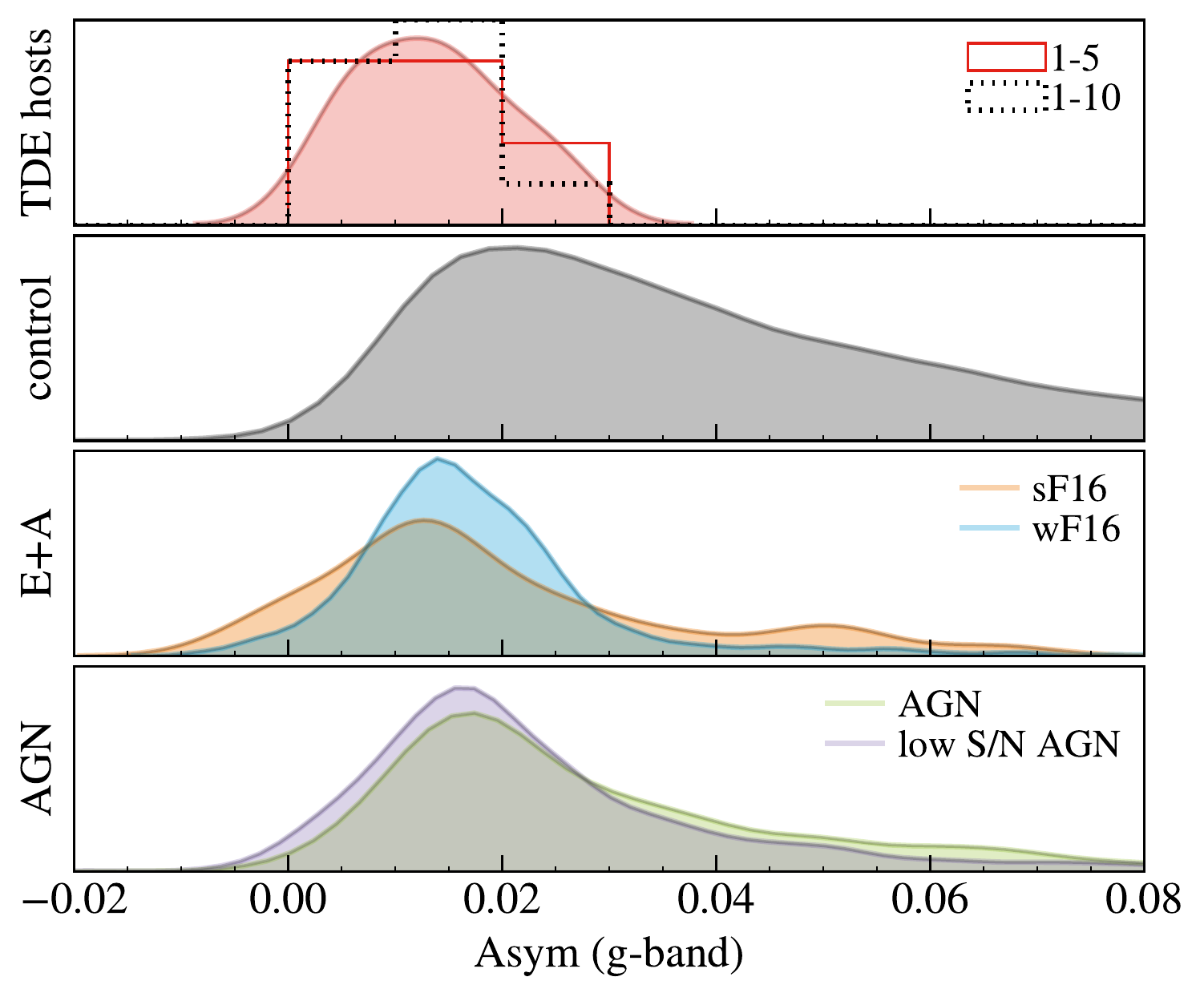}
\caption{
Left panel: galaxy asymmetry indicators in the $g$ and $r$ bands for TDE host galaxies and our reference catalog. Middle panel: asymmetry indicator in the $g$ band vs. BH mass. The $r$ band is similar. BH masses for TDE hosts 8, 9, and 10 are determined via $M_{\star, {\rm bulge}}$. Contours are spaced by $0.5\sigma$, with the darkest shading containing $0.5\sigma$ and the lightest shading containing $2\sigma$. Average errors in the TDE host galaxy measurements are shown in the top left (asymmetry indicators do not have associated errors in our catalog). Right panel: asymmetry indicator distribution in the $g$ band in different subsamples, matched on BH mass of TDE hosts 1-5. The $r$ band is similar. 1D histograms are smoothed (we also show the true histogram for the TDE hosts 1-5 in red and 1-10 in dotted black) and normalized to equal area.
}
\label{fig:Asym}
\end{figure*}

\subsection{Merger Indicators}
Galaxy mergers might also enhance the TDE rate if they trigger binary BH inspiraling. To study this, we use the {\it RA}1\_2 galaxy asymmetry indicators output from GIM2D \citep[as defined in][]{Simard:2002, Simard:2009} in the $g$ and $r$ bands. For recent major mergers, the asymmetry indicators can be $\gtrsim 0.04$ \citep[e.g.,][]{Patton:2016}. The asymmetry indicators are shown plotted against each other as well as against BH mass for TDE host galaxies and our reference catalog in Figure~\ref{fig:Asym}. In the right two panels, we show only $g$-band measurements; results are similar for $r$ band. Interestingly, our TDE host galaxies have small asymmetry indicators, suggesting that they are not the products of recent major mergers. It is important to note that a small SDSS asymmetry indicator does not necessarily correspond to the lack of a merger. The major limitation here is the SDSS resolution, along with the fact that asymmetries in mergers tend only to be high for major mergers and gas-rich galaxies \citep{Lotz:2010, Ji:2014}. In the right panel (controlled for BH mass), we see that TDE host galaxies and quiescent Balmer-strong galaxies both have a narrower distribution in asymmetry indicator compared to the reference catalog. They do not share the reference catalog's tail toward high asymmetry indicators.

\subsection{Summary of TDE Host Galaxy Properties; SDSS Images}
\begin{figure*}[tbp]
\epsscale{1.15}
\plotone{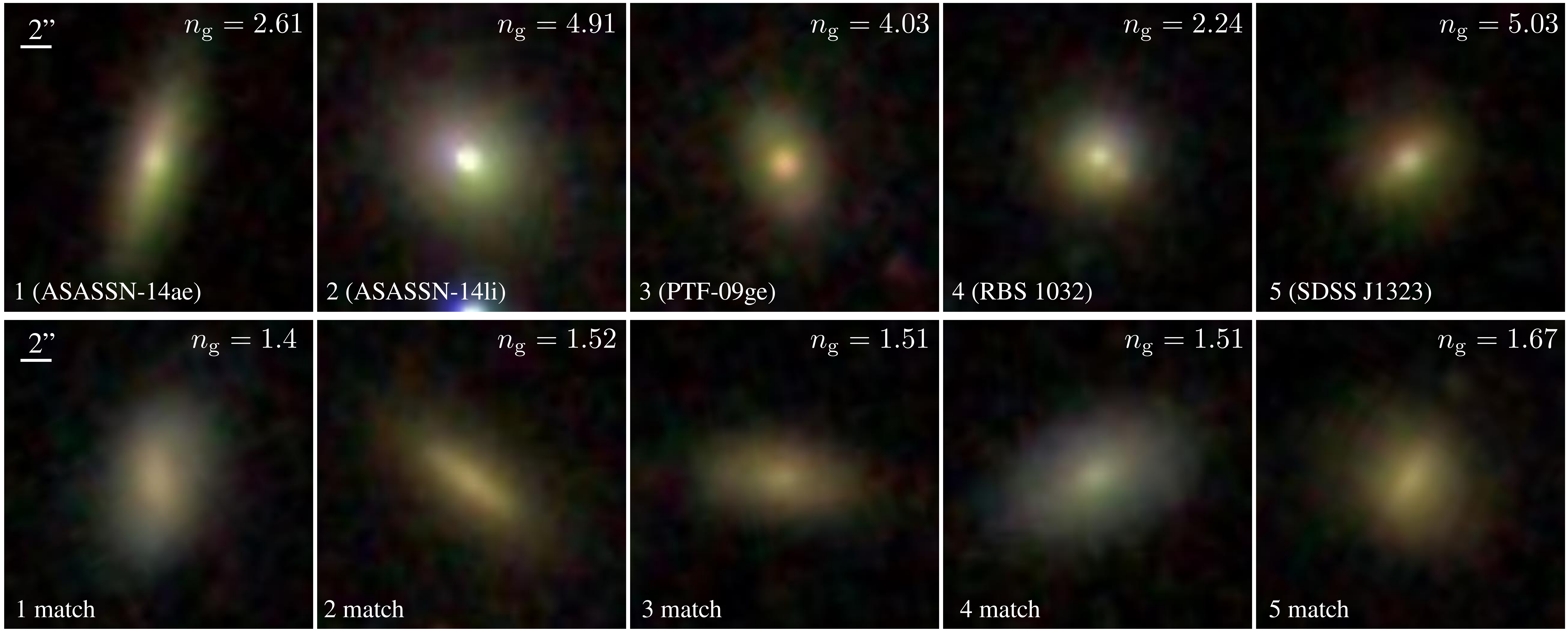}
\caption{
Top panels: SDSS $gri$ images of TDE host galaxies 1-5. Bottom panels: for each TDE host galaxy, a randomly selected galaxy matched in BH mass and redshift to the TDE host galaxy, but with a galaxy \Sersic index, bulge $g-r$, and half-light surface brightness very close to the median values of our reference catalog at that BH mass. Images are $20\arcsec \times 20\arcsec$. The galaxy \Sersic index of each galaxy is listed in the top right of each image.
}
\label{fig:sdss}
\end{figure*}

In summary, TDE host galaxies tend to have bluer bulges, lower half-light surface brightnesses, and more centrally concentrated light profiles (in \Sersic index and bulge fraction) than ``typical'' galaxies at their BH masses. As an illustrative example of this, in Figure~\ref{fig:sdss}, we show SDSS images\footnote{Though higher resolution imaging is available for several of these galaxies, we show SDSS photometry here as this is what was used to derive the photometric galaxy properties in our reference catalog.} of TDE host galaxies 1-5 as well as, for each TDE host galaxy, a randomly selected galaxy matched in BH mass and redshift to the TDE host galaxy, but with a galaxy \Sersic index, bulge $g-r$, and half-light surface brightness very close to the median values of our reference catalog at that BH mass. The galaxies in the bottom panels are thus ``typical'' galaxies in a few of the parameters we considered above, but matched in BH mass and redshift to the TDE host galaxies. The higher central concentration of these TDE host galaxies is visually apparent. The galaxy \Sersic index of each galaxy is listed in the top right of each image.

%%%%%%%%%%%%%%%%%%%%%%%%%%%%%%%%%%%%%%%%%%%%%%%%%%
\section{DISCUSSION}\label{sec:discuss}
In this section, we discuss (1) whether there is a selection effect against detecting TDEs in SF galaxies, (2) if the observed time delay between SF and AGN activity can help us understand the nature of the post-starburst TDE delay, (3) if the quiescent Balmer-strong enhancement can be understood in terms of \Sersic indices in the green valley, and (4) the implications of higher central light concentrations on the TDE rate. Before discussing these issues, we list a summary of our key findings.

\subsection{Summary}
We studied TDE host galaxies in the context of a catalog of $\sim$500,000 SDSS galaxies. Our main conclusions are:
\begin{enumerate}

\item Controlling for (by creating matched samples) selection effects due to BH mass, redshift completeness, bulge color, and half-light surface brightness reduces the apparent overrepresentation of TDEs in E+A host galaxies by a factor of $\sim$4
(from $\sim$$\times$100-190 to $\sim$$\times$25-48 in sF16 galaxies), but cannot fully explain the preference.

\item Controlling for BH mass, TDE host galaxies have bluer bulge $g-r$ colors (by $\sim$0.3~mag) and fainter half-light surface brightnesses (by $\sim$1~mag/arcsec$^2$) than galaxies in our reference catalog. TDE host galaxies have low galaxy asymmetry indicators, suggesting that they are not the result of a recent major merger.

\item TDE hosts and E+A galaxies have high galaxy \Sersic indices and high $B/T$ for their BH masses, suggesting a higher stellar density in their cores. On average, our TDE host galaxies have galaxy \Sersic indices and bulge fractions in the top 10\% of those of reference catalog galaxies at their BH masses. We identify a region in galaxy \Sersic index versus BH mass space that contains $\sim$2\% of our reference catalog galaxies but 5/5 (or 6/10) of our TDE host galaxies. 
\end{enumerate}

We also note that \citet{Graur:2017} appeared on arXiv after submission of this work and is an independent and complementary analysis of TDE host galaxy properties. They study a smaller set of host galaxy properties, but, importantly, find that TDE host galaxies have high stellar surface mass densities. This is similar to our finding that TDE host galaxies appear more centrally concentrated, with higher galaxy \Sersic indices and $B/T$. They control for the type of galaxy in which the TDE is found (quiescent or star-forming) and find that this result is driven particularly by the star-forming hosts.

\subsection{Is There a Selection Effect Against Detecting TDEs in SF Galaxies?}
In order to understand the TDE rate enhancement in quiescent Balmer-strong galaxies, it is important to know whether there is a selection effect against detecting TDEs in SF galaxies. As SF galaxies are (by definition) not quiescent Balmer-strong galaxies, if there is a bias against observing TDEs in SF galaxies, quiescent Balmer-strong galaxies are less rare as hosts.

First, we consider whether the dust and gas associated with SF may obscure TDEs. Certainly, starbursting galaxies have significant dust attenuation \citep[e.g.,][]{Casey:2014}, especially in the optical/UV bands where the SF occurs. \citet{Del-Moro:2013} study a robust sample of 51 ``hidden" radio-excess AGNs in SF galaxies, and find that half of these are not detected in deep {\it Chandra} X-ray data, indicating that they might be heavily obscured. As a case study, the Seyfert 2 galaxy NGC 4968 is found to have heavy obscuration and circumnuclear SF, as well as SF-associated gas that may increase the covering factor of the enshrouding gas and play a role in obscuring the AGN \citep{LaMassa:2017}. If a TDE occurred in NGC 4968, the large column density ($N_{\rm H} > 1.25 \times 10^{24}\ \mathrm{cm^{-2}}$) would prevent X-ray---and optical/UV, for dust-to-gas ratios similar to the Milky Way---identification.
 
Although difficult to detect, AGNs are often found in SF galaxies\footnote{Though we caution that this strongly depends on how the AGN is selected \citep[see e.g.,][]{Ellison:2016}.} \citep{Bongiorno:2012, Ellison:2016}. Nuclear activity in SF galaxies might be fairly common, and it is possible that TDEs are missed in these galaxies primarily due to selection effects. We note that \citet{Tadhunter:2017} discovered a TDE in a nearby ultraluminous infrared galaxy.\footnote{This galaxy is not in our sample as it is not in the SDSS catalogs we draw from.} The galaxy features suggest that there is an unusually clear view of the nuclear star-forming region, whose obscuration is known to have a complex structure \citep[e.g.,][]{Buchner:2017}. Two galaxies in our TDE host galaxy sample, SDSS J0748 and SDSS J1342 \citep{Wang:2012}, are in SF galaxies. We note, however, that both events were classified as TDEs in a search for extreme coronal line emitters, and in both cases were not identified based on their light curve properties.  

Second, we recognize that there could be a bias against TDE identification in SF galaxies, as TDE characterization for nuclear transients is not done systematically. In addition, current TDE host galaxies have relatively low half-light surface brightnesses, while SF galaxies have relatively high half-light surface brightnesses (see Figure~\ref{fig:scatters}). If image subtraction is less robust for these galaxies, this could help explain the lack of TDEs in SF galaxies.

\subsection{Time Delay between SF and AGN/TDEs}
If TDEs occur preferentially in post-starburst galaxies, then the TDE rate is time dependent and, in particular, depends on the recent SF history of the galaxy. If the starburst was caused by a galaxy merger, the second inspiraling BH could certainly enhance the TDE rate; we discuss this further below, but first we consider the relationship between the SF and the TDE. After a starburst occurs, the gas is transported inwards on some timescale and may drive an enhanced TDE rate \citep[possibly through contributing to disk instabilities;][]{Madigan:2017}. 

This transport timescale is also seen in AGN activity, and there is an intriguing connection between AGN activity and SF episodes. AGN activity appears to be triggered by the same gas that drives SF episodes \citep{Trump:2015}, but with a time delay, presumably due to the transport timescale of gas to the BH. The post-starburst timescale in TDE hosts appears tantalizingly similar to this observed AGN activity--SF episode delay.

\citet{Wild:2010} study a sample of 400 galaxies with BH masses of $10^{6.5}$-$10^{7.5}M_\sun$ that have experienced a starburst in the past 600 Myr. They find that the average rate of accretion of matter onto the BH rises steeply $\Approx 250$ Myr after the starburst begins. Similarly, \citet{Davies:2007} study the nuclei of nine AGNs at spatial scales of $\Approx10$ pc and find a hint of a delay of 50-100 Myr between the onset of star formation and accretion onto the BH. This delay is strikingly reminiscent of the post-starburst timescale inferred for TDE hosts, which have post-starburst ages of 10-1000 Myr \citep{French:2017}\footnote{Though note that only 4/8 of their sample have post-starburst ages of $<$250 Myr.}.

\citet{LaMassa:2013} study the connection between AGN activity and star formation with a sample of $\Approx$ 28,000 obscured active galaxies. They find that circumnuclear star formation is associated with increased BH activity and that angular momentum transfer through the disk limits the efficiency of mass inflow onto the BH.
\citet{Mullaney:2012} also suggest that the same secular processes that drive the bulk of star formation are responsible for the majority of SMBH activity, which gives further credence to the idea that the majority of moderate nuclear activity is fueled by internal mechanisms rather than violent mergers \citep{Mullaney:2012a}. If the connection between SF history and the TDE rate is analogous to the SF episode--AGN activity connection, this could help explain the quiescent Balmer-strong galaxy preference.

\subsection{The Green Valley, \Sersic Index, and E+A/Post-Starburst Galaxies}
\begin{figure*}[tbp]
\epsscale{0.6}
\plotone{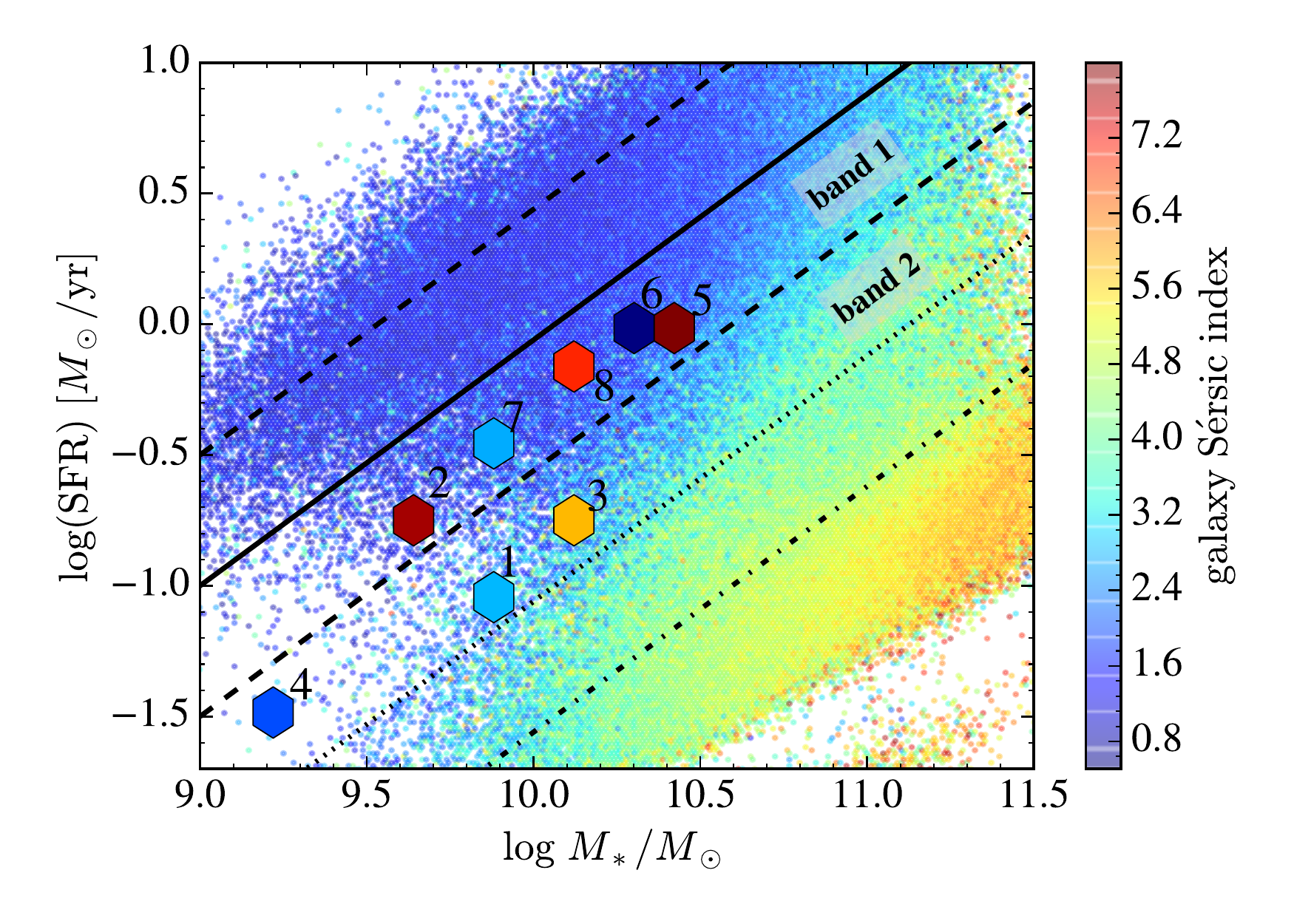}
\epsscale{0.5}
\plotone{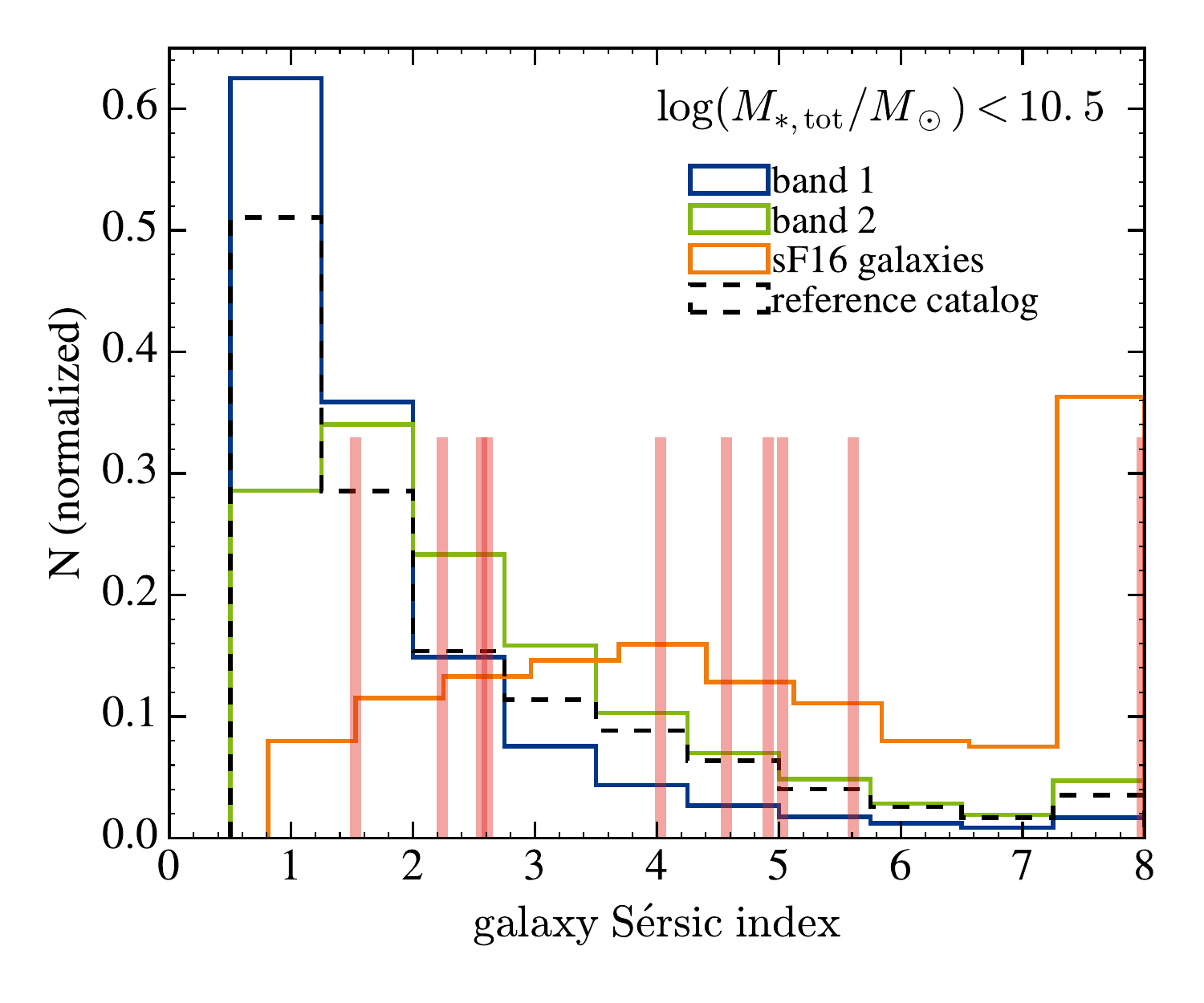}
\caption{
Left panel: total star formation rate vs. total stellar mass for our reference catalog and TDE host galaxies 1-8. Color corresponds to galaxy \Sersic index, ranging from 0.5 (blue) to 8 (red); for the reference catalog galaxies, this is the mean within each hexagonal bin.
Right panel: normalized histograms of galaxy \Sersic index for our reference catalog (dashed black), galaxies between the SFMS and 0.5~dex below the SFMS (band 1; blue), galaxies between the SFMS $-$ 0.5~dex and the SFMS $-$ 1.0~dex (band 2; green), and galaxies in the sF16 selection (orange). \Sersic indices of our TDE hosts are indicated by the red vertical lines. Here, we restrict all samples to $\log(M_{\star,{\rm tot}}/M_\sun)<10.5$ to roughly match the range of TDE host galaxy values.
}
\label{fig:sfr2}
\end{figure*}

In Figure~\ref{fig:sfr}, we saw that our TDE host galaxies lie below the SFMS, but not by more than 1.0~dex (some are in the green valley and some appear near it), and that E+A/post-starburst galaxies inhabit a similar region.
In the left panel of Figure~\ref{fig:sfr2}, we show SFR versus $M_{\ast, {\rm tot}}$ for our TDE host galaxies and the reference catalog, similarly to Figure~\ref{fig:sfr}, but color-coded by galaxy \Sersic index; this shows the evolution of galaxy surface density profiles in this parameter space. TDE host galaxies clearly inhabit a transition region in \Sersic index. 

In the right panel of Figure~\ref{fig:sfr2}, we show normalized histograms of \Sersic index for different subsamples of the reference catalog: the band from the SFMS to 0.5~dex below the SFMS (band 1), the band 0.5~dex below this, from the dashed to the dotted line (band 2), and galaxies in the sF16 selection. Though their population is small, E+A galaxies have much higher \Sersic indices relative to the larger population of galaxies in the green valley.
If we restrict our reference catalog to galaxies with $\log(M_{\star,{\rm tot}}/M_\sun) < 10.5$, to match the TDE hosts, as well as to galaxy \Sersic indices of $>2.0$, the percentage of sF16 (wF16) galaxies in band 2 is 1.9\% (16\%). Further restricting to galaxies with \Sersic indices $>4.0$ results in 3.8\% sF16 galaxies and 23\% wF16 galaxies in band 2. This cut also results in 4.9\% sF16 galaxies in band 1.
So, E+A galaxies are a subset of green valley galaxies with high \Sersic indices.
If, due to their higher intrinsic rates in these galaxies, TDEs are preferentially found in post-starburst galaxies with high central densities, then E+A/post-starburst galaxies are relatively less rare as TDE hosts.

\subsection{On the TDE Rate Enhancement}
The increased TDE rates in E+A/post-starburst galaxies might be explained by mergers and/or by higher nuclear stellar densities.

A merger is thought to increase the disruption rate by many orders of magnitude \citep[e.g.,][]{Ivanov:2005}.  But these rate enhancements are short lived (less than 1 Myr), so that the fraction of TDEs resulting from merging BHs is expected to be  low \citep{Wegg:2011}. The asymmetry indicators for TDE hosts are comparatively small, meaning that they show no obvious signs of recent major mergers. However, TDE hosts could be the product of high mass ratio mergers. Indeed, unequal mass ratio mergers are more effective at enhancing the TDE rate as stars are scattered into the loss cone of disruptive orbits more efficiently \citep{Chen:2009}. \citet{French:2017} find that the post-starburst ages of TDE hosts, if the starburst arose from a galaxy merger, are consistent with mergers of mass ratios more equal than 12:1 for most hosts, which is still consistent with our findings.

Additionally, higher resolution observations could reveal signs of mergers that the asymmetry indicators miss. Using MUSE integral field unit (IFU) spectroscopy observations, \citet{Prieto:2016} find that the host galaxy of ASASSN-14li shows asymmetric and filamentary structures---signs of a recent merger---yet this galaxy has a small {\it RA}1\_2 asymmetry indicator (0.008 and 0.015 in the $g$ and $r$ bands, respectively).

Higher \Sersic indices and bulge-to-total-light ratios for both TDE host galaxies and E+A/post-starburst galaxies (see Figure \ref{fig:sersic_bt}) provide a natural explanation for the enhanced disruption rates in these galaxies. \citet{Stone:2016a} were the first to predict that the enhanced rate in E+A/post-starburst galaxies might be due to their large central stellar densities, as per-galaxy TDE rates scale roughly as $\dot N_\text{TDE} \propto \rho_\star^2$. We caution that our galaxy \Sersic indices were derived using measurements that typically do not resolve the nuclear regions of the galaxy and, as such, the  density of the sphere of influence of the BH cannot be directly constrained. 
We note, however, that \citet{Simard:2011} isolate a subsample of $\sim$53,000 galaxies with justified free-$n$ \Sersic fits to their bulges, and they find that galaxies with low and high $n_{\rm bulge}$ values also have low and high $n_{\rm g}$ values.

There is also some direct evidence that E+A galaxies have higher central stellar densities. \citet{Stone:2016} find that the E+A galaxy NGC 3156 is centrally overdense, leading to an estimated TDE rate via two-body relaxation of $\Sim10^{-3}$ yr$^{-1}$, an order of magnitude higher than for other galaxies with similar BH masses. \citet{Pracy:2012} study a sample of seven local E+A galaxies with IFU spectroscopy and find that they have compact young cores and stellar population gradients that are predicted from models of mergers and tidal interactions that funnel gas into the galaxy core. This suggests that these galaxies are being seen in the late stage of a merger where the nuclei have already coalesced.

Importantly---and separate from understanding the E+A galaxy preference---we are able to identify a photometric criterion (light concentration, given by either the \Sersic index or bulge fraction) that may predict a TDE overabundance more broadly than a spectroscopic criterion (E+A classification). For upcoming transient surveys where there may be up to $\sim$10$^{5}$ transients discovered each year, a photometric host galaxy selection criterion could be extremely useful for focusing limited follow-up resources.  For instance, choosing nuclear transients in high-\Sersic galaxies could significantly increase the success of confirming TDEs.

\acknowledgements
We thank Trevor Mendel for useful conversations, as well as for performing bulge+disk decompositions on two TDE host galaxies (numbers 9 and 10) that are in the \citet{Simard:2011} catalog but not in the \citet{Mendel:2014} catalog.
We thank Jarle Brinchmann, Viraj Pandya, Peter Behroozi, Kevin Bundy, Xavier Prochaska, and Andrea Merloni  for useful conversations. J.L.-S. and E.R.-R. are grateful for support from the Packard Foundation, NASA ATP grant NNX14AH37G and NSF grant AST-1615881. S.L.E. acknowledges the receipt of an NSERC Discovery Grant. The UCSC transients group is supported in part by NSF grant AST-1518052 and from fellowships from the Alfred P. Sloan Foundation and the David and Lucile Packard Foundation to R.J.F.

\software{corner \citep{Foreman-Mackey:2017}, Matplotlib \citep{http://dx.doi.org/10.1109/MCSE.2007.55}, seaborn \citep{https://doi.org/10.5281/zenodo.824567}.}

%%%%%%%%%%%%%%%%%%%%%%%%%%%%%%%%%%%%%%%%%%%%%%%%%%
\appendix

\section{Other TDE Host Galaxy Properties}\label{sec:other}

\begin{figure*}[tbp]
\epsscale{0.56}
\plotone{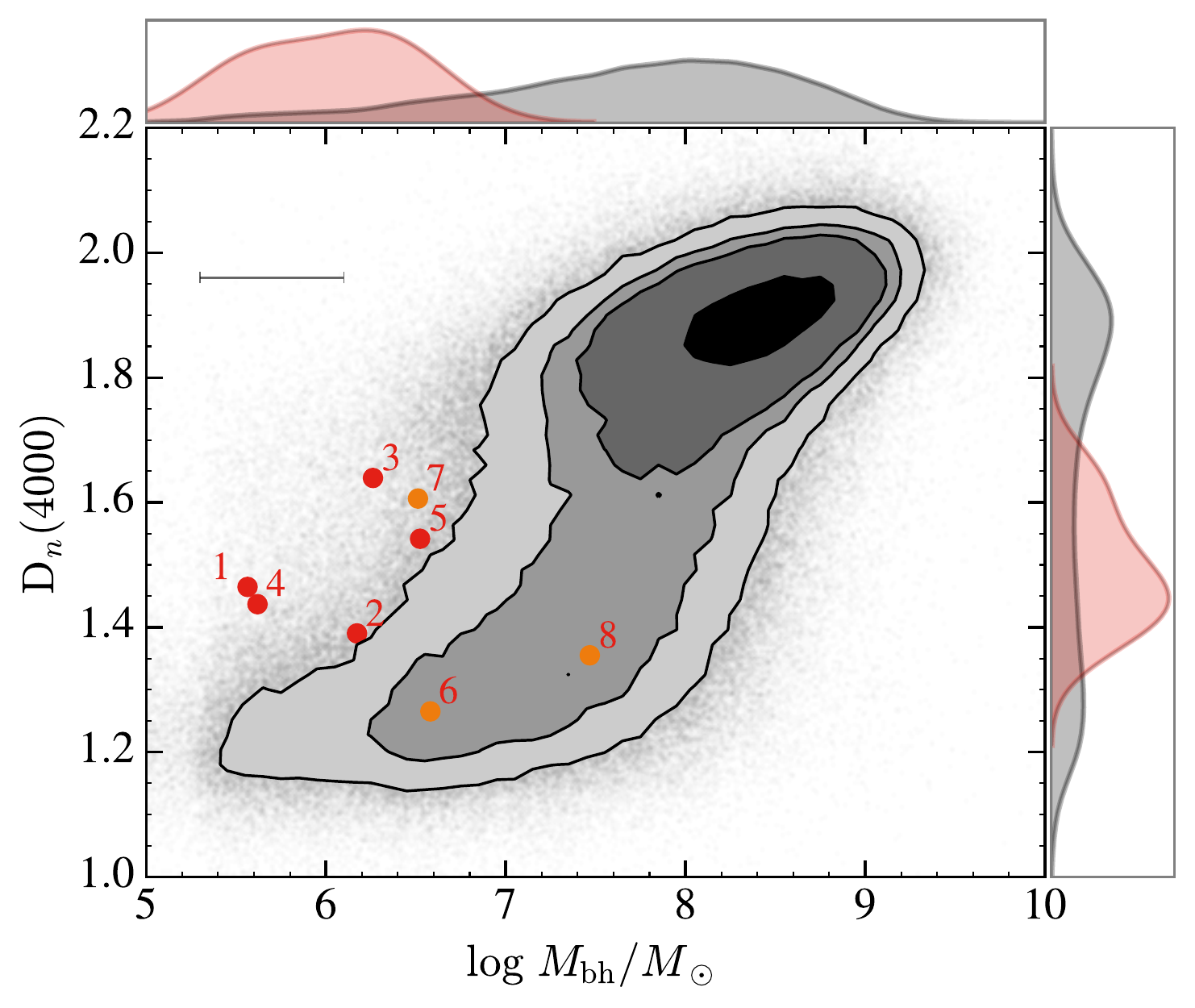}
\plotone{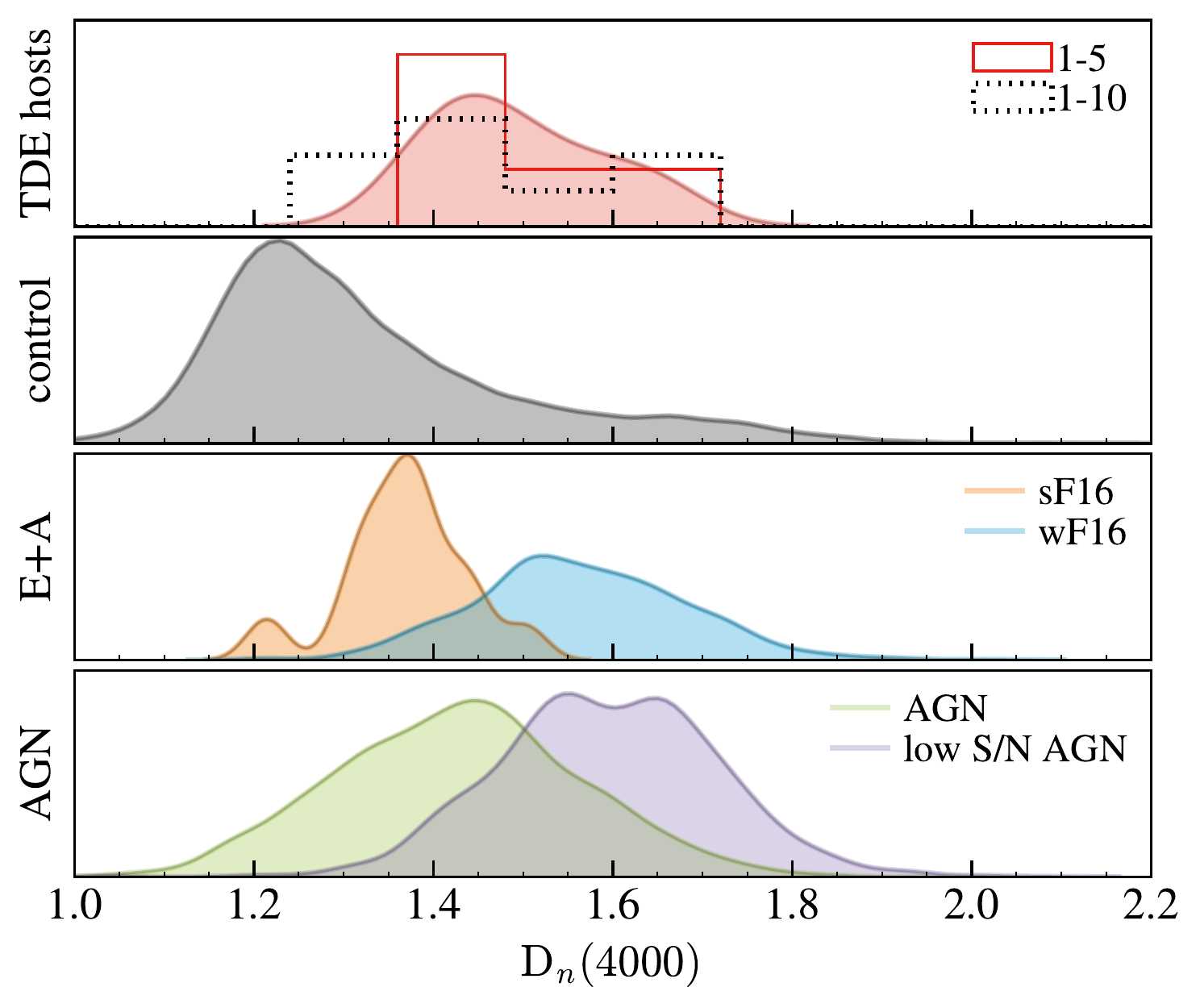}
\plotone{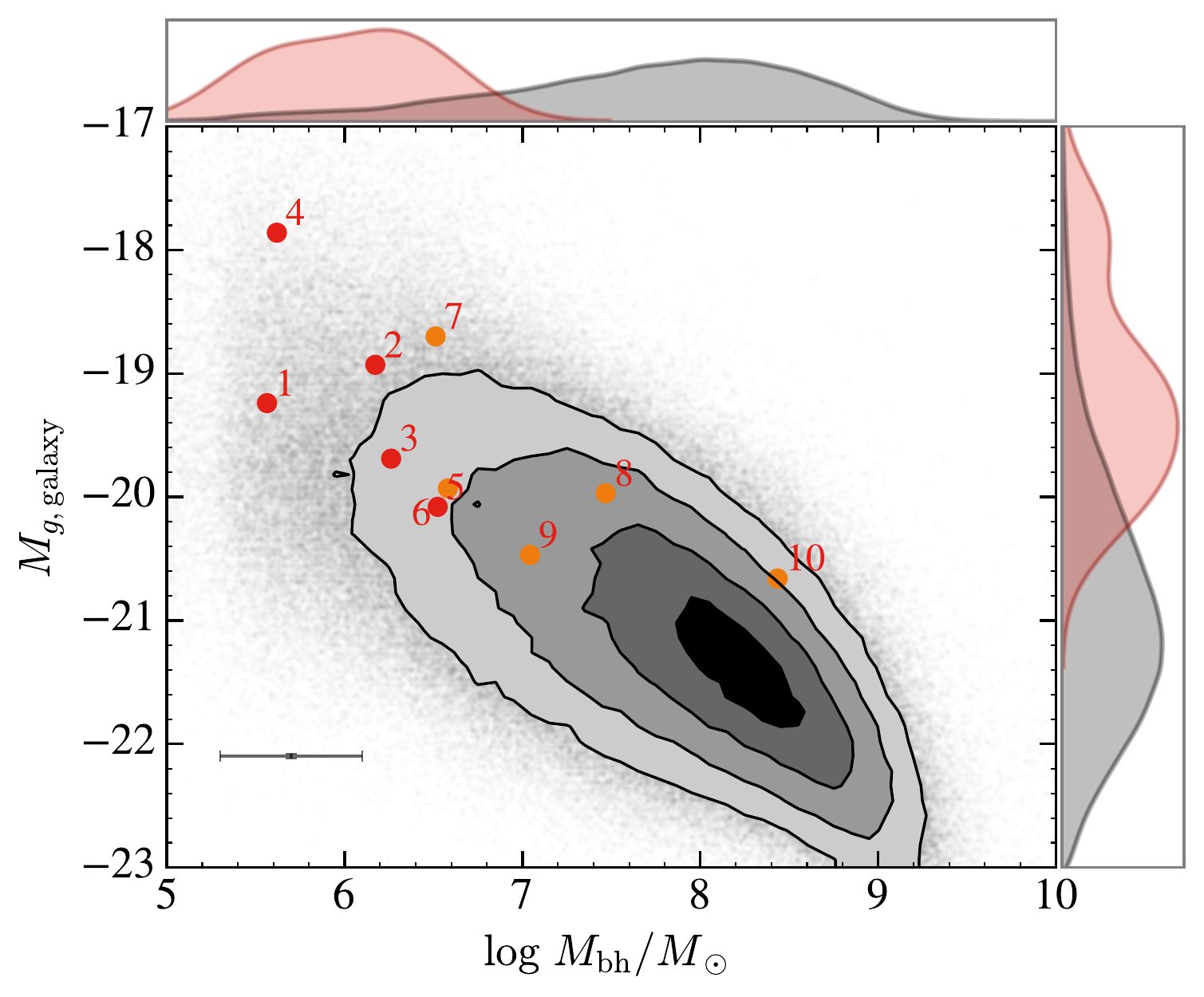}
\plotone{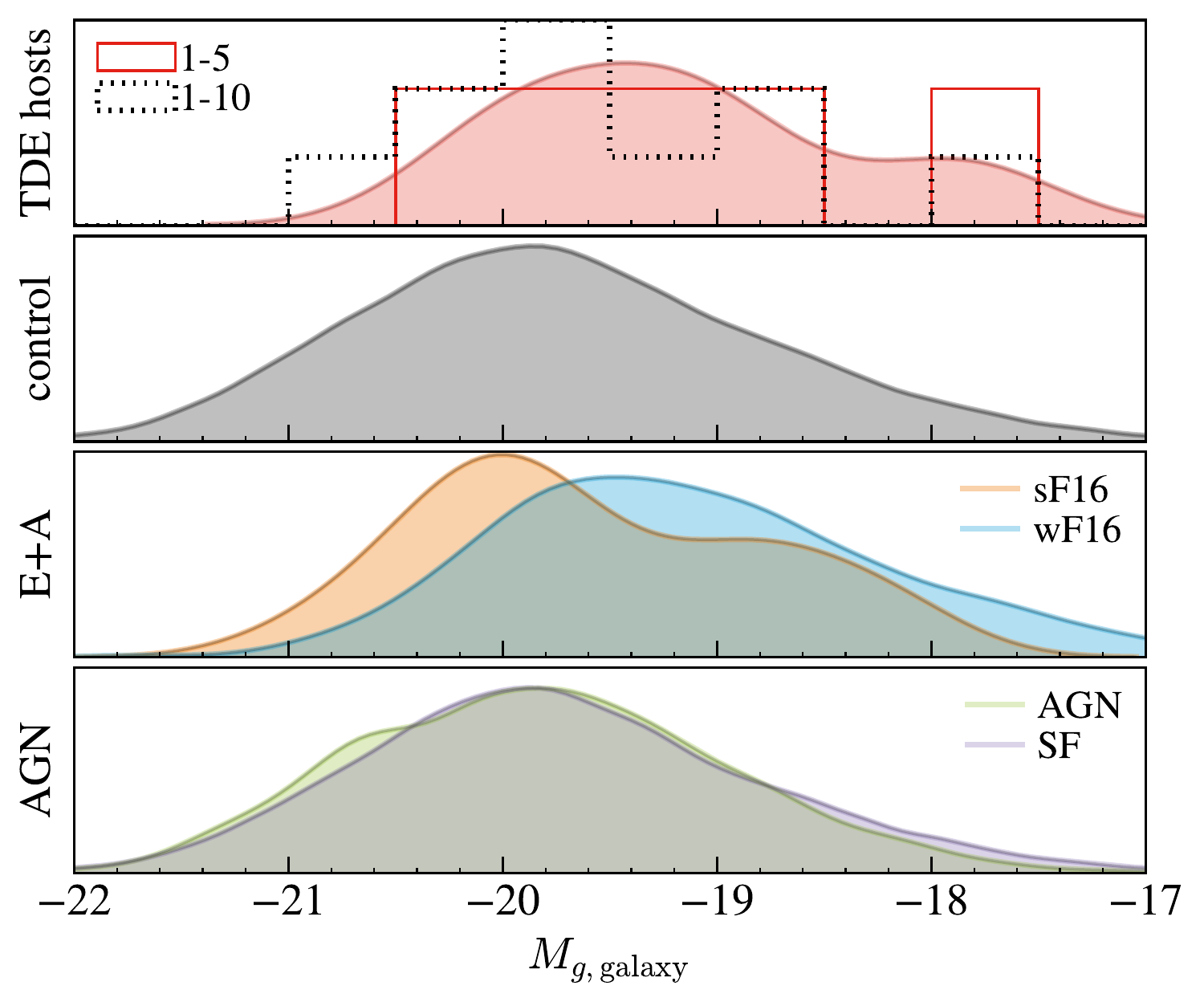}
\plotone{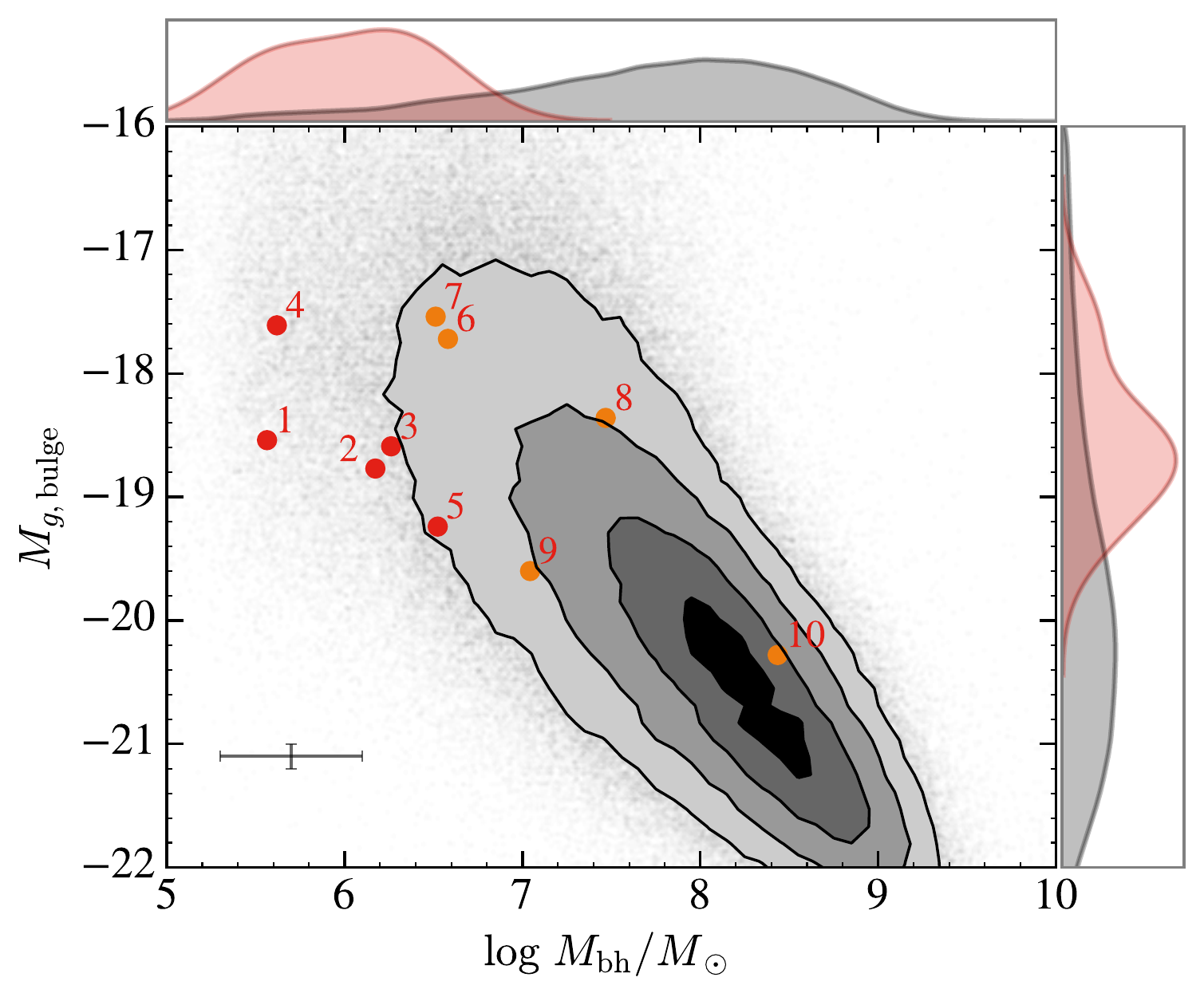}
\plotone{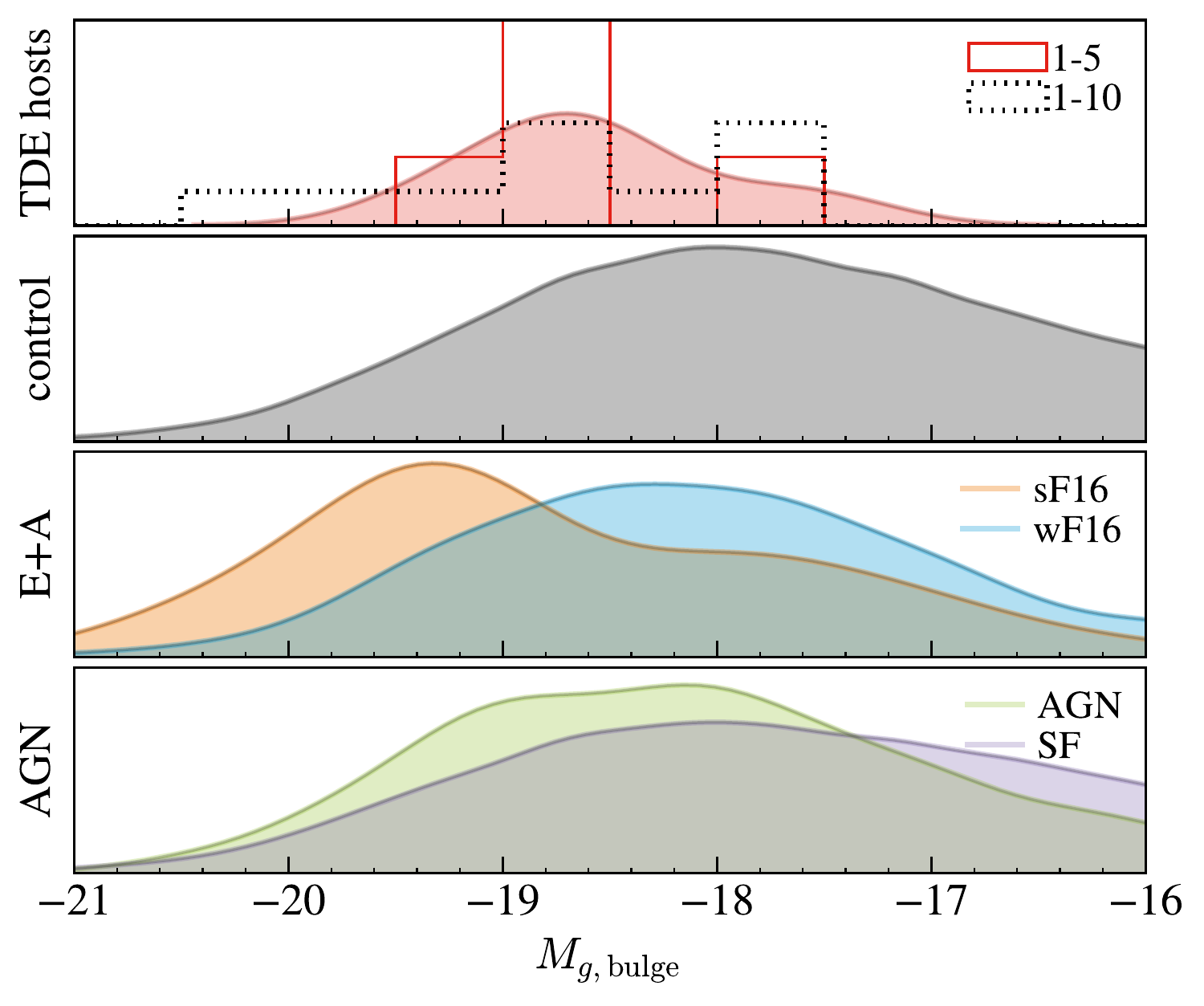}
\caption{
Left panels, top to bottom: D$_n$(4000), $g$-band galaxy absolute magnitude, and $g$-band bulge absolute magnitude vs. BH mass for TDE host galaxies and our reference catalog. For galaxy and bulge magnitudes, the results are similar for the $r$-band. BH masses for TDE hosts 8, 9, and 10 are determined via $M_{\star, {\rm bulge}}$. Contours are spaced by $0.5\sigma$, with the darkest shading containing $0.5\sigma$ and the lightest shading containing $2\sigma$. Average errors in the TDE host galaxy measurements are shown in the top or bottom left. Right panels: 1D distributions of these properties in different subsamples, matched on BH mass of TDE hosts 1-5. All 1D distributions are smoothed and normalized to equal area. Unsmoothed 1D histogram for TDE hosts 1-5 is shown in solid red, and for TDE hosts 1-10 in dotted black.
}
\label{fig:other1}
\end{figure*}

\begin{figure*}[tbp]
\epsscale{0.56}
\plotone{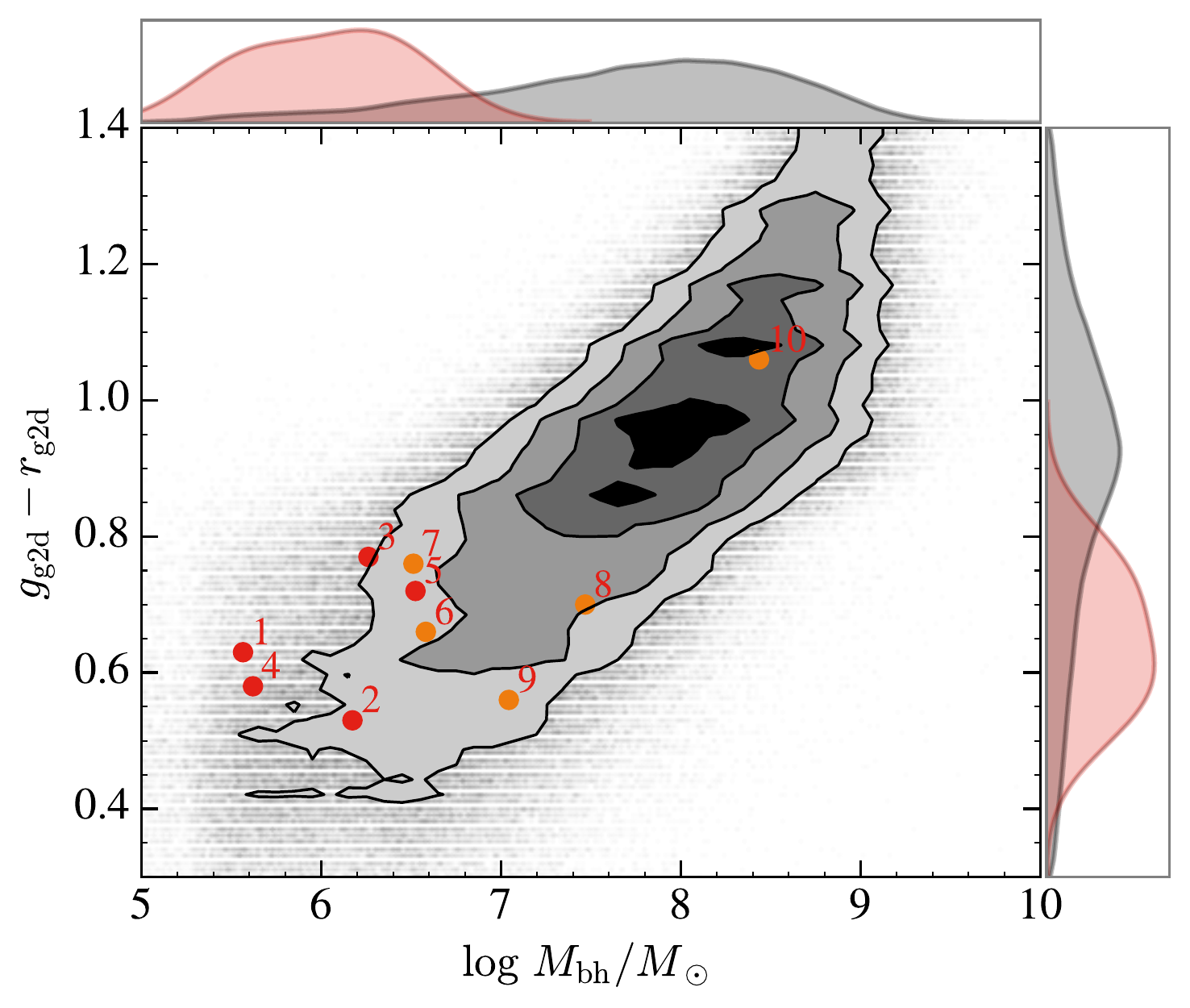}
\plotone{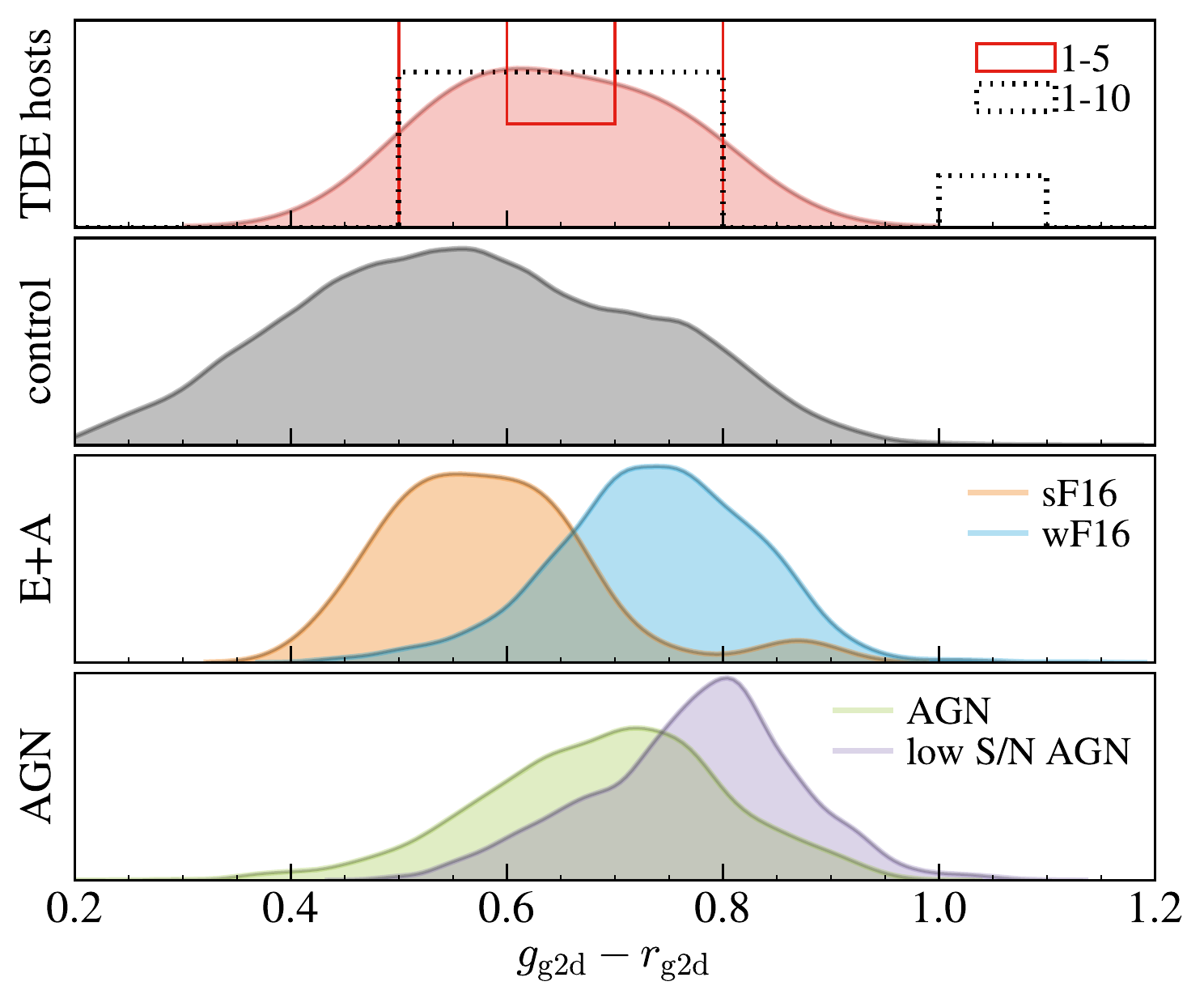}
\plotone{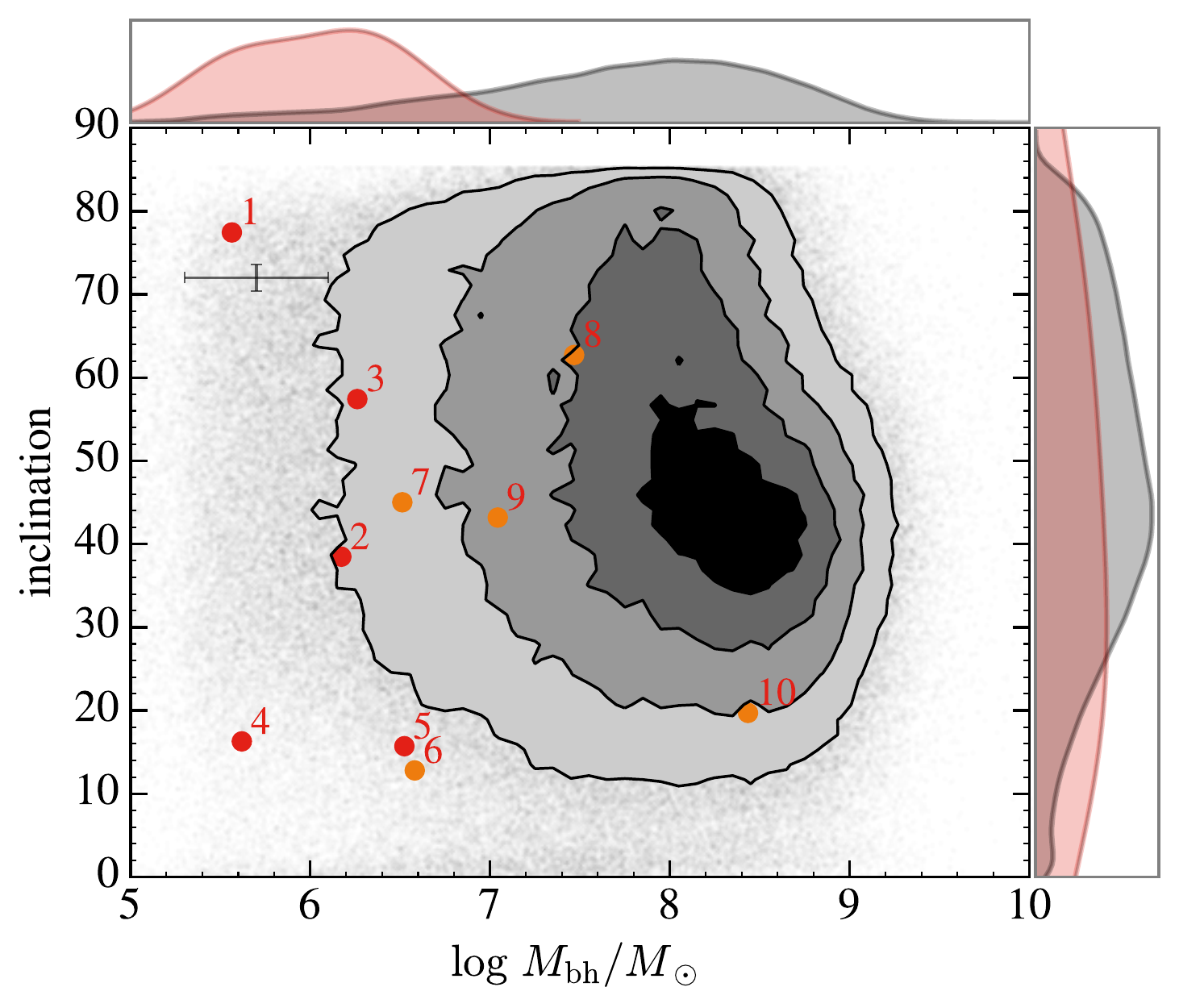}
\plotone{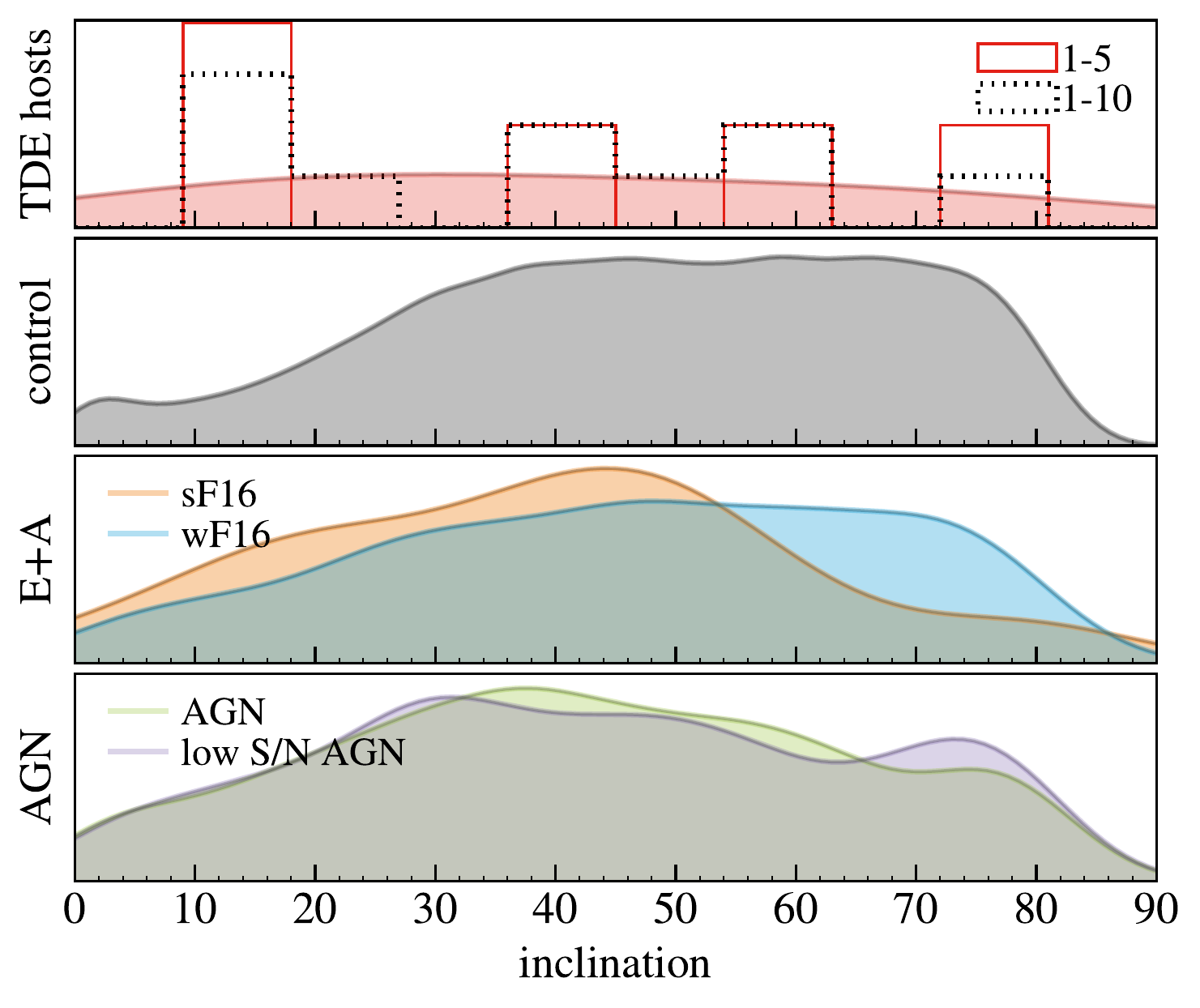}
\caption{
Same description as in Figure \ref{fig:other1}, but galaxy $g-r$ in the top panel and inclination (face-on is $0^\circ$, maximum of 85$^\circ$) in the bottom panel.
}
\label{fig:other2}
\end{figure*}

Here, we discuss a few other properties of the host galaxies of TDEs. The top-left panel of Figure \ref{fig:other1} shows D$_n$(4000) versus BH mass for TDE host galaxies and our reference catalog. D$_n$(4000) corresponds to the strength of the 4000\AA~break and is an indicator of the age of the galaxy stellar population: no break means a young galaxy and a strong break means an old galaxy. Following \citet{Kauffmann:2003b, Kauffmann:2003a} and \citet{Brinchmann:2004}, the peak in D$_n$(4000) at $\Sim1.3$ corresponds to galaxies with $r$-band weighted mean stellar ages of $\Sim1$-3 Gyr and mass-weighted mean ages a factor of $\Sim2$ larger. The peak in D$_n$(4000) at $\Sim1.85$ corresponds to older elliptical galaxies with mean stellar ages of $\Sim10$ Gyr. The TDE host galaxies lie roughly in between these two peaks, with a peak in D$_n$(4000) at $\Sim1.5$, indicating that they have mean stellar ages in between those of these two populations. Note that this measurement is only sensitive to the dominant stellar population and does not reveal multiple stellar populations. In the right panel, where we match on BH mass to TDE hosts 1-5, we see that sF16 galaxies exhibit a younger mean stellar age than wF16 galaxies. AGNs exhibit a younger mean stellar age than low-S/N AGNs. We note that D$_n$(4000) is not as effective a metric as the H$\alpha$ EW and Lick H$\delta_A$ metric (Figure \ref{fig:french}) in isolating quiescent Balmer-strong galaxies.

The middle-left panel of Figure \ref{fig:other1} shows $g$-band galaxy absolute magnitude versus BH mass for TDE host galaxies and our reference catalog. Results are similar for $r$ band. Controlling for BH mass (right panel), TDE hosts are slightly fainter than the catalog galaxies, and sF16 galaxies are slightly brighter than wF16 galaxies. The AGN and SF samples have very similar distributions. The bottom panel of Figure \ref{fig:other1} shows bulge $g$-band absolute magnitude. Results are similar for $r$ band. Here, controlling for BH mass, both TDE host galaxies and (to a somewhat greater extent) sF16 galaxies have brighter bulge magnitudes than the reference catalog. Medians and spreads on the distributions of the galaxy and bulge absolute magnitudes are given in Table \ref{tab:medians}.

Lastly, we show galaxy $g-r$ in the top panel of Figure \ref{fig:other2}, and galaxy inclination (face-on is $0^\circ$) in the bottom panel. Controlling for BH mass, the galaxy $g-r$ colors for TDE host galaxies and our reference catalog appear fairly similar. sF16 galaxies have bluer colors than wF16 galaxies. The inclinations of TDE hosts appear fairly uniform, with a hint of a preference for lower inclinations. Interestingly, the inclination distribution of sF16 galaxies appears different from that of the reference catalog, with a preference for $20 \lesssim i \lesssim 65$ in contrast to the catalog's preference for $40 \lesssim i \lesssim 80$. Medians and spreads on the distributions of galaxy colors and inclinations are given in Table~\ref{tab:medians}.

\section{Correlations}\label{sec:cor}
\begin{figure*}[tbp]
\epsscale{1.22}
\plotone{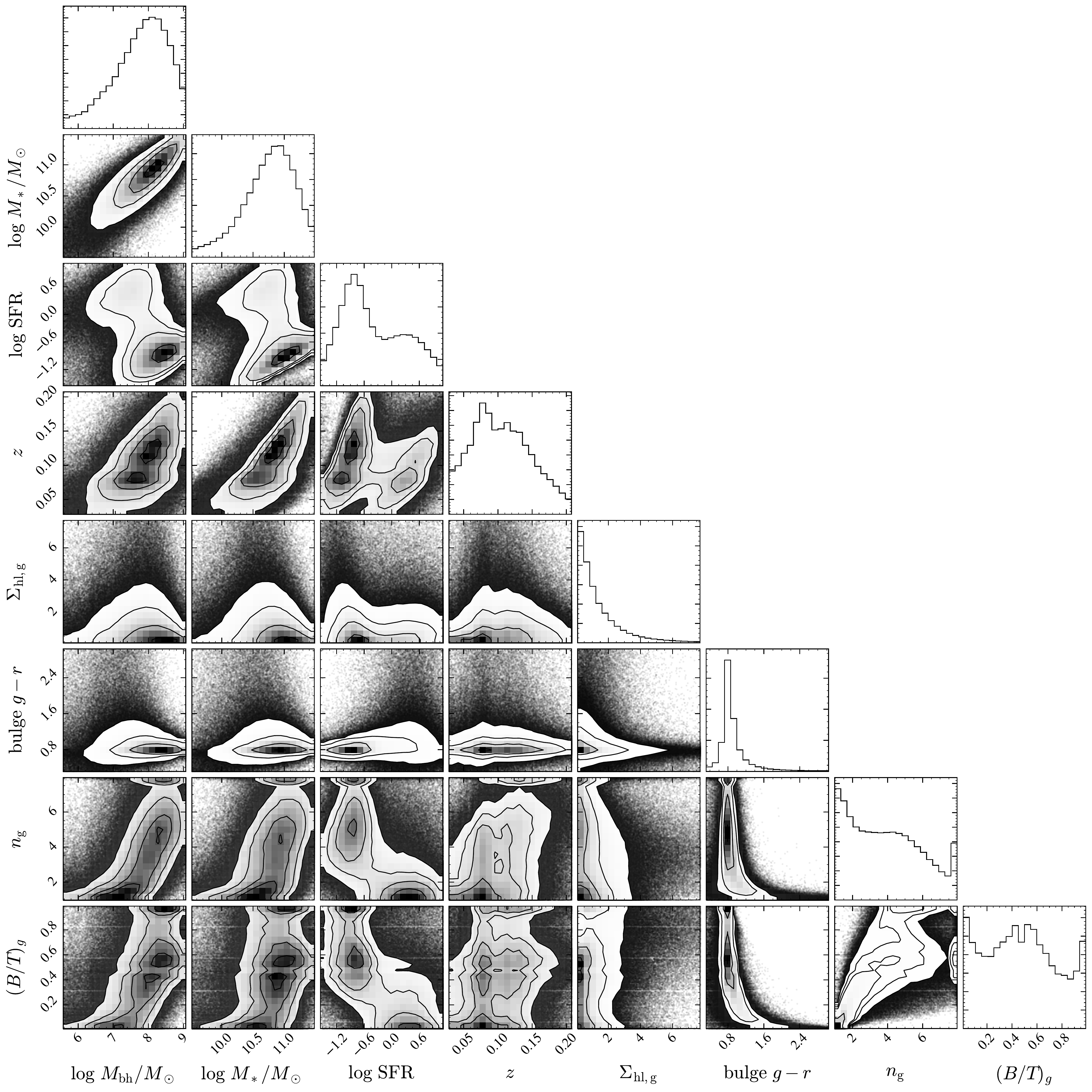}
\caption{
Correlations between many of the properties explored in this paper for our reference catalog of $\sim$500,000 galaxies. From left to right along the bottom row, the properties are BH mass, total stellar mass, total star formation rate (in $M_\odot$ yr$^{-1}$), redshift, $g$-band half-light surface brightness (in mag/arcsec$^2$), bulge $g-r$, galaxy \Sersic index ($n_{\rm g}$), and $g$-band bulge-to-total-light ratio $(B/T)_g$. Each panel contains 95\% of the points.
}
\label{fig:corner}
\end{figure*}

We show correlations between many of the properties explored in this paper in Figure~\ref{fig:corner}. We show BH mass, total stellar mass, redshift, half-light surface brightness, bulge $g-r$, galaxy \Sersic index, and bulge-to-total-light ratio. The total stellar mass behaves very similarly to BH mass in this diagram, which, in addition to it being a more physically relevant parameter for tidal disruptions, is why we use BH mass in our analysis. However, as mentioned in the text, if we simply replace BH mass with $M_{\star, {\rm tot}}$ in our analysis, our conclusions remain the same.

\bibliography{refs}
\bibliographystyle{aasjournal}

\end{document}